\documentclass[10pt,aps,pra,twocolumn,superscriptaddress,floatfix,longbibliography]{revtex4-1}

\usepackage{graphicx}
\usepackage{amsfonts}
\usepackage{amssymb}
\usepackage{amsmath}
\usepackage{txfonts}
\usepackage{lipsum}
\usepackage{color}
\usepackage{wasysym}
\usepackage[colorlinks=true, allcolors=blue]{hyperref}
\usepackage{bbold}
\usepackage{etoolbox} 
\usepackage[normalem]{ulem}
\usepackage{mathtools} 
\usepackage{multirow} 
\usepackage{hhline}
\usepackage{mathrsfs} 
\usepackage{tabularx}

\usepackage[dvipsnames]{xcolor}
\usepackage{tikz}
\usetikzlibrary{arrows}
\usetikzlibrary{shapes.geometric}
\usetikzlibrary{decorations}
\usetikzlibrary{decorations.pathmorphing}
\usetikzlibrary{decorations.markings}

\tikzset{
interaction/.style={
  decorate,
  line cap=round,
  decoration={snake}
},
interactionBold/.style={
  double,
  line cap=round,
  decorate,
  decoration={snake}
},
electron/.style={
  line cap=round,
  postaction={ decorate },
  decoration={ 
    markings, 
    mark=at position .5 with {
      \node[isosceles triangle, fill=black, transform shape,
      scale=0.4, anchor=center] {};
   }
  }
},
electronBold/.style={double, line cap=round,
  postaction={ decorate },
  decoration={ 
    markings, 
    mark=at position .5 with {
      \node[isosceles triangle, fill=black, transform shape,
      scale=0.4, anchor=center] {};
   }
  }
},
TriangleVertex/.style={
  shape=isosceles triangle,
  isosceles triangle apex angle=60,
  inner sep=0pt,
  minimum height=0.4cm,
  fill=Green!50,
  draw,
  anchor=apex,
},
SmallTriangleVertex/.style={
  shape=isosceles triangle,
  isosceles triangle apex angle=60,
  inner sep=0pt,
  minimum height=0.3cm,
  fill=Green!50,
  draw,
  anchor=apex,
},
}

\newcommand\orangesout{\bgroup\markoverwith{\textcolor{orange}{\rule[0.5ex]{2pt}{0.4pt}}}\ULon}

\usepackage{ulem}

\usepackage{comment}

\begin{document}

\title{Electron-magnon dynamics triggered by an ultrashort laser pulse: A real-time Dual $GW$ study} 

\author{Nagamalleswararao Dasari}
\email{nagamalleswararao.d@gmail.com}
\affiliation{Institut f{\"u}r Theoretische Physik, Universit{\"a}t Hamburg, Notkestra{\ss}e 9, 22607 Hamburg, Germany}

\author{Hugo U. R. Strand}
\affiliation{School of Science and Technology, {\"O}rebro University, SE-70182 {\"O}rebro, Sweden}

\author{Martin Eckstein}
\affiliation{Institut f{\"u}r Theoretische Physik, Universit{\"a}t Hamburg, Notkestra{\ss}e 9, 22607 Hamburg, Germany}
\affiliation{The Hamburg Centre for Ultrafast Imaging, Luruper Chaussee 149, 22761 Hamburg, Germany}

\author{Alexander I. Lichtenstein}
\affiliation{Institut f{\"u}r Theoretische Physik, Universit{\"a}t Hamburg, Notkestra{\ss}e 9, 22607 Hamburg, Germany}
\affiliation{European X-Ray Free-Electron Laser Facility, Holzkoppel 4, 22869 Schenefeld, Germany}
\affiliation{The Hamburg Centre for Ultrafast Imaging, Luruper Chaussee 149, 22761 Hamburg, Germany}

\author{Evgeny A. Stepanov}
\affiliation{CPHT, CNRS, {\'E}cole polytechnique, Institut Polytechnique de Paris, 91120 Palaiseau, France}
\affiliation{Coll\`ege de France, Universit\'e PSL, 11 place Marcelin Berthelot, 75005 Paris, France}

\begin{abstract}

Ultrafast irradiation of correlated electronic systems triggers complex dynamics involving quasiparticle excitations, doublons, charge carriers, and spin fluctuations. To describe these effects, we develop an efficient non-equilibrium approach, dubbed \mbox{D-$GW$}, that enables a self-consistent treatment of local correlations within dynamical mean-field theory (DMFT) and spatial charge and spin fluctuations, that are accounted for simultaneously within a diagrammatic framework. 
The method is formulated in the real-time domain and provides direct access to single- and two-particle momentum- and energy-dependent response functions without the need for analytical continuation, which is required in Matsubara frequency-based approaches. We apply the \mbox{D-$GW$} method to investigate the dynamics of a photo-excited extended Hubbard model, the minimal system that simultaneously hosts strong charge and spin fluctuations.
Focusing on the challenging parameter regime near the Mott transition, we demonstrate that correlated metals and narrow-gap Mott insulators undergo distinct thermalization processes involving complex energy transfer between single-particle and collective electronic excitations.
\end{abstract}

\maketitle

\section{Introduction}

Light-induced ultrafast phenomena, such as superconductivity~\cite{Claudio2016, Cavalleri2018}, hidden quantum states~\cite{ichikawa2011, Stojchevska2014}, dielectric breakdown~\cite{yamakawa2017, Diener2018}, and ultrafast melting~\cite{Iwai2003, Afanasiev2019}, have attracted much attention in condensed matter research due to their potential technological applications~\cite{basov2017}. 
Understanding these non-thermal phenomena requires addressing the real-time evolution of a quantum many-body system, which remains an outstanding challenge for both experiment and theory. This task is highly demanding due to the non-perturbative nature of the electronic correlations responsible for these effects and the wide range of relevant time scales, spanning from a few femtoseconds to tens of picoseconds.

The availability of ultrashort light pulses with a wide frequency range has enabled the rapid development of novel experimental tools to characterize photo-excited states~\cite{Torre2021, Boschini2024}. 
However, theoretical progress in this field still lags behind experimental advancements due to the absence of accurate real-time numerical methods capable of simulating large systems, accessing long time scales, and incorporating all relevant correlation effects. 

The {\it state-of-the-art} method for solving correlated time-dependent electronic problems is the dynamical mean-field theory (DMFT)~\cite{Georges1996}.
This method accounts for the effect of local electronic correlations by mapping the original lattice model onto an effective impurity problem.
In the late 2000s, this approach, called non-equilibrium DMFT~\cite{Aoki2014}, was formulated on the L-shaped Keldysh contour to describe the real-time dynamics of correlated systems. 
It incorporates the time-dependence of the electronic correlations through the local self-energy $\Sigma(t,t')$, which is a function of two-time argument, rather than just a time difference $(t'-t)$ as it would be in equilibrium.
The two-time dependence of $\Sigma(t,t')$ is essential for the temporal evolution of correlated systems, which is highly challenging to calculate using exact real-time impurity solvers, for example, quantum Monte Carlo and matrix-product states~\cite{Eckstein2010, Gramsch2013, Blazer2015, wolf2014}.
The former methods suffer from a sign problem, and the real-time evolution is limited to a short time scale. The latter does not have a sign problem but is restricted to one-dimensional systems. 
For this reason, in actual calculations $\Sigma(t,t')$ is obtained by solving the time-dependent impurity problem using {\it state-of-the-art} perturbative methods~\cite{Dasari2018, Mott-NCA}, such as iterative perturbation theory (IPT), non-crossing approximation (NCA), and one-crossing approximations (OCA).
Over a decade, non-equilibrium DMFT explained some of the dynamical phenomena observed in most of the correlated materials
\cite{Yuta2023}
, for example, the emergence of non-thermal states \cite{li2018}, ultrafast melting of collective orders \cite{Afanasiev2019}, non-thermal quantum criticality \cite{Tsuji2013}, dielectric breakdown of Mott-insulators \cite{Ecsktein20102}, prethermalization \cite{Eckstein2009}, ultrafast dynamics of doublons and charge \cite{Ligges2018,Petocchi2023,PhysRevB.102.220301}
At the same time, the DMFT assumption of the locality of the self-energy fails for the 
systems with predominant non-local collective electronic fluctuations.
These fluctuations give rise to various collective instabilities in the quantum systems, such as superconductivity or charge and spin orders, and are thus pivotal in understanding the physics of correlated systems.
The real-time evolution of these fluctuations is the prime research interest, because competing thermodynamic instabilities are the key to stabilising hidden quantum states with novel orders that are otherwise impossible in equilibrium. 

Effects of the non-local fluctuations in the weakly correlated systems have been studied using the 
fluctuation-exchange (FLEX)~\cite{BICKERS1989206, PhysRevB.57.6884, Dasari20181, Dasari2019}, $GW$~\cite{GW1, GW2, GW3}, and two-particle self-consistent (TPSC)~\cite{PhysRevB.55.3870, refId0, Simard2022} methods.
In the strongly correlated systems, where non-perturbative effects are significant, the non-local fluctuations have been treated via the cluster~\cite{PhysRevB.58.R7475, PhysRevB.62.R9283, PhysRevLett.87.186401, Maier2005, doi:10.1063/1.2199446, RevModPhys.78.865, PhysRevLett.101.186403, PhysRevB.94.125133, Cluster_Fratino} or diagrammatic~\cite{Rohringer2018, Lyakhova_review} extensions of DMFT. 
The former considers a finite cluster of lattice sites instead of a single-site DMFT impurity problem, which allows for the exact treatment of spatial correlations within the considered cluster. 
The main challenge in these methods is solving the cluster problem, which scales exponentially with cluster size and the number of orbitals. 
Therefore, cluster theories can only handle short-range electronic correlation in practice~\cite{PhysRevLett.101.256404, PhysRevB.89.195146, PhysRevB.91.235107}. 
On the other hand, diagrammatic extensions use the DMFT impurity problem as a reference system, and correlation effects beyond the reference system are incorporated by considering a certain subset of Feynman diagrams.  

To date, the $GW$~\cite{PhysRevLett.90.086402, PhysRevLett.92.196402, PhysRevLett.109.226401, PhysRevB.87.125149, PhysRevB.90.195114, PhysRevB.94.201106, PhysRevB.95.245130} and TPSC~\cite{PhysRevB.107.075158, PhysRevB.107.235101} extensions of DMFT are the most advanced real-time diagrammatic methods that have been implemented on the 
Keldysh contour.
The non-equilibrium $GW$+EDMFT approach~\cite{PhysRevLett.118.246402, PhysRevB.100.041111, PhysRevB.100.235117} has been introduced to account for the non-local charge fluctuations essential for understanding photo-induced screening effects. 
This method also incorporates the spatial Coulomb interaction but, at the same time, neglects the non-local spin fluctuations that dominate most low-dimensional correlated materials. 
The TPSC+DMFT approach~\cite{PhysRevB.107.245137} is the most recent attempt to incorporate both charge and spin fluctuations out of equilibrium. 
However, this method does not account for non-local Coulomb interactions and is, in practice, limited to small local Coulomb interaction strengths.

The simultaneous incorporation of spatial charge and spin fluctuations in the presence of non-local~\cite{stepanov2021coexisting, vandelli2024doping, stepanov2024signatures}, and even frequency-dependent~\cite{stepanov2021coexisting, stepanov2023charge}, electronic interactions has been achieved in the recently developed equilibrium method known as the dual triply irreducible local expansion (\mbox{D-TRILEX}) approach~\cite{PhysRevB.100.205115, PhysRevB.103.245123, 10.21468/SciPostPhys.13.2.036}, which was inspired by the TRILEX method~\cite{PhysRevB.92.115109, PhysRevB.93.235124, PhysRevB.96.104504, PhysRevLett.119.166401}. \mbox{D-TRILEX} is a diagrammatic extension of DMFT that allows for a self-consistent description of local electronic correlations and the leading spatial collective fluctuations of an arbitrary range~\cite{stepanov2021coexisting, vandelli2024doping, PhysRevLett.132.236504, PhysRevB.110.L161106, stepanov2024signatures, Cuprates}, which is an advantage over cluster theories.
This approach enables calculation of both single- and two-particle quantities, with accuracy comparable to significantly more complex diagrammatic methods~\cite{PhysRevB.103.245123, 10.21468/SciPostPhys.13.2.036, PhD_Vandelli}. Furthermore, \mbox{D-TRILEX} has a rather simple $GW$-like diagrammatic structure, which enables efficient calculations in the multi-band framework~\cite{PhysRevLett.127.207205, PhysRevResearch.5.L022016, PhysRevLett.129.096404, PhysRevLett.132.226501, stepanov2023charge, Ruthenates}.

The success of the multi-band implementation motivated us to formulate the time-dependent version of the \mbox{D-TRILEX} method. 
It is important to mention that the equilibrium \mbox{D-TRILEX} method is formulated in the Matsubara frequency space and accounts for the exact three-point (Hedin~\cite{GW1}) vertex corrections of the reference (usually DMFT impurity) problem in the diagrams for the self-energy and the polarization operator. Calculating the three-point vertex out-of-equilibrium involves enormous computational resources and requires developing real-time impurity solvers to calculate this three-time-dependent object. 
In addition, computing the electronic self-energy and the polarization operator on an L-shaped Keldysh contour with these vertices involves four-time integration and amounts to the generalisation of standard Langereth rules to the much more general situations, which is extremely challenging \cite{Kugler2021}. Unfortunately, with the available numerical tools, these complications prevent the efficient implementation of the time-dependent generalization of the \mbox{D-TRILEX} approach, making it prohibitively expensive computationally.

In this work, we introduce a ``lighter'' version of the \mbox{D-TRILEX} method on the L-shaped Keldysh contour by approximating the three-point vertex function by its instantaneous (short-time limit) component.
With this approximation, the diagrammatic structure of the proposed method resembles that of the $GW$+DMFT theory, which inherently accounts for charge fluctuations while additionally incorporating spatial magnetic excitations on an equal footing.
To avoid confusion among the readers, we call this method Dual-$GW$ (\mbox{D-$GW$}) throughout the paper.
  
To showcase the effectiveness of our developed method, we apply it to analyze the non-equilibrium dynamics of the single-band extended Hubbard model, focusing on the most challenging regime, i.e., near the Mott transition, on both its metallic and insulating sides.
Our simulations reveal that paramagnetic metals reach a thermal state within the simulation time, while Mott insulators do not, despite the same excitation protocol.
In Mott insulators, the pre-thermal state is characterized by photo-excited doublons and non-local charge and spin fluctuations, which reach a temperature indicative of a unique thermal state. 
This is different from that of quasi-particles. 
True thermalization occurs only when quasi-particles in the transient also achieve this unique thermal state temperature, which our simulations could not fully reach.
An intriguing finding is that even with antiferromagnetic spin fluctuations, impact ionization- a nonlinear relaxation process for photo-excited charge carriers- remains highly favorable for extended systems. In this process, the time evolution of the occupied density of states illustrates the transfer of excess kinetic energy from the high-energy photo-excited doublons (holons) to the underlying antiferromagnetic spin fluctuations. This energy transfer can be directly measured using time-resolved photoemission spectroscopy.

The paper is organized as follows. 
Section~\ref{sec:model-method} discusses the model and method, focusing on developing the necessary formalism and numerical tools for solving \mbox{D-$GW$}. 
Section~\ref{sec:equilibrium} examines the \mbox{D-$GW$} equilibrium phase diagram of the extended Hubbard model. In Section~\ref{sec:photo-doping}, we present the photo-excited dynamics of the extended Hubbard model near the metal-to-Mott-insulator transition triggered by a short electric pulse. Finally, we summarize our findings in Section~\ref{sec:conclusion}.

\section{Model and Method \label{sec:model-method}}

\subsection{Model}

\begin{figure}[t!]
\includegraphics[width=0.68\linewidth]{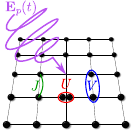}
\caption{The extended Hubbard model [Eq.~(\ref{eq:model})] on the square lattice with nearest neighbor hopping $J$, local Hubbard interaction $U$, and nearest-neighbor interaction $V$, in a time-dependent external electric field ${\mathbf{E}}_{p}(t)$.}
\label{fig:model}
\end{figure}

The canonical model for strongly correlated electrons that possesses strong collective charge and spin fluctuations is the half-filled extended Hubbard model on the square lattice
\begin{equation}
\mathcal{H} = 
J\sum_{\langle ij \rangle, \sigma} c^{\dagger}_{i\sigma}c^{\phantom{\dagger}}_{j\sigma}
+ U \sum_i n^{\phantom{\dagger}}_{i\uparrow} n^{\phantom{\dagger}}_{i\downarrow} 
+ V\sum_{\langle ij \rangle,\sigma\sigma'} n^{\phantom{\dagger}}_{i\sigma} n^{\phantom{\dagger}}_{j\sigma'}, 
\label{eq:model}
\end{equation}
where $c^{(\dagger)}_{i\sigma}$ annihilates (creates) an electron with spin ${\sigma \in \left\{ \uparrow,\downarrow \right\}}$ on the lattice site $i$. 
$J$ is the hopping amplitude between the nearest-neighbor sites ${\langle ij \rangle}$, $U$ is the on-site Coulomb potential between the electronic densities ${n_{i\sigma} = c^\dagger_{i\sigma} c^{\phantom{\dagger}}_{i\sigma}}$, and $V$ is the short-range Coulomb interaction between the neighboring sites. 
The sketch of the model is shown in Fig.~\ref{fig:model}.
At half-filling the hopping $J$ gives perfect Fermi surface nesting and conspires with the local Hubbard interaction $U$ inducing antiferromagnetic (AFM) correlations.
The nearest-neighbor interaction $V$, on the other hand, drives charge fluctuations that can induce charge density wave order of holes and doublons. 

The electron coupling to the electrical field $\mathbf{E}_{p}(t)$ of light is introduced through the Peierls substitution. 
In terms of the single-particle dispersion ${\varepsilon(\mathbf{k}) =-2J \left(\cos k_x + \cos k_y \right)}$, the induced vector potential $\mathbf{A}(t) = - \int_0^t \mathbf{E}_{p}(\bar{t}) \, d\bar{t}$, enters as a time-dependent momentum shift in the single particle dispersion ${\varepsilon\left(\mathbf{k},t\right) \equiv \varepsilon\left(\mathbf{k} - \frac{1}{\hbar} \mathbf{A}(t)\right)}$. We take the hopping $J$ as the unit of energy. We further set electric charge $e$, speed of light $c$ and lattice constant $a$ to unity ($e$ = $c$ = $a$ = 1) throughout the paper. This choice defines the unit of time $\hbar/J$. 

\subsection{D-TRILEX method:}

\begin{figure}[t]
\includegraphics[width=1\linewidth]{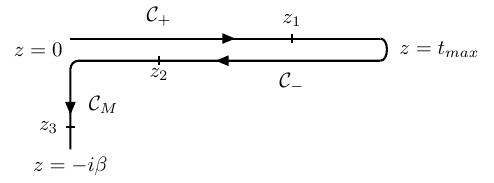}
\caption{Non-equilibrium time contour $\mathcal{C}$ in the complex time plane, consisting of the forward ($\mathcal{C}_+$), backward ($\mathcal{C}_-$) and imaginary ($\mathcal{C}_M$) time branches, $\mathcal{C} = \mathcal{C}_+ \cup \mathcal{C}_- \cup \mathcal{C}_M$, with general contour times $z_1$, $z_2$, $z_3 \in \mathcal{C}$.}\label{fig:contour}
\end{figure}

Applying an external time-dependent perturbation to the interacting electronic problem~\eqref{eq:model} triggers complex dynamics of interplaying single-particle and collective electronic excitations.  
Following the time evolution of such a system requires using an advanced non-equilibrium many-body approach that can describe all these excitations simultaneously. In this work, we present a real-time implementation of the \mbox{D-$GW$} method, a ``lighter'' version of the (\mbox{D-TRILEX}) approach, that is capable of self-consistent treatment of single- and two-particle fluctuations in the charge and spin channels with no restriction on their range. To this, we began with grand-canonical partition function $\mathcal{Z}$ = $\mathrm{Tr}\left[T_{\mathcal{C}} e^{i\cal{S}}\right]$, where $T_{\mathcal{C}}$ denotes the contour-ordering operator on the Kadanoff-Baym contour $\mathcal{C}$ in the complex time plane, see Fig.~\ref{fig:contour}. 
The contour times ${z \in \mathcal{C}}$ runs from ${t=0}$ to the maximum simulation time $t_{max}$ along the real-time forward branch $\mathcal{C}_+$, back to ${t=0}$ along the backward branch $\mathcal{C}_-$, and then to $-i\beta$ along the imaginary-time branch $\mathcal{C}_M$, for details see \cite{Stefanucci:2013oq}. 
The partition function in the coherent-state path-integral formalism can be expressed as ${\mathcal{Z} = \int D[c^{*},c] \exp(i{\cal S})}$, where
\begin{align}
{\cal S} = & -\int_{\cal C} dz \, \Bigg\{- \sum_{ij,\sigma} c^{*}_{i\sigma}(z) \left[\delta^{\phantom{*}}_{ij}(i\partial^{\phantom{*}}_{z}+\mu) - J^{\phantom{*}}_{ij}(z)\right] c_{j\sigma}^{\phantom{*}}(z) \nonumber \\  
& + \sum_{i} U n_{i\uparrow}(z) \, n_{i\downarrow}(z) + \frac{1}{2} \int_{\cal C} dz' \sum_{ij,\varsigma} \rho^{\varsigma}_{i}(z) V^{\varsigma}_{ij} \, \rho^{\varsigma}_{j}(z') \Bigg\}
\label{eq:action_latt}
\end{align}
is the action corresponding to the Hamiltonian~\eqref{eq:model}.
We will focus on the single-orbital case with general hopping integral and non-local interactions, but multi-orbital systems can be
approached similarly.  
$V^{\varsigma}_{ij}$ refers to the non-local interaction in the charge (${\varsigma={\rm d}}$) and
spin (${\varsigma={\rm m}\in \{x, y, z \}}$) channels. 
The variables $\rho^{\varsigma}_{i}(z) = n^{\varsigma}_{i}(z) - \langle n^{\varsigma}_{i} \rangle$ describe fluctuations of the charge and
spin densities $n^{\varsigma}_{i}(z)=\sum_{\sigma\sigma'}c^{*}_{i\sigma}(z) \sigma^{\varsigma}_{\sigma\sigma'} c^{\phantom{*}}_{i\sigma'}(z)$ around
their average values. ${\sigma^{\rm d} = \mathbb{1}}$, and $\sigma^{\rm m}$ are the Pauli matrices in the spin space.

\mbox{D-TRILEX}~\cite{PhysRevB.100.205115, PhysRevB.103.245123, 10.21468/SciPostPhys.13.2.036} belongs to a family of ``dual'' approaches~\cite{PhysRevB.77.033101, PhysRevB.79.045133, PhysRevLett.102.206401, PhysRevB.94.035102, PhysRevB.96.035152, BRENER2020168310, Rubtsov20121320, PhysRevB.90.235135, PhysRevB.93.045107, PhysRevB.94.205110, Stepanov18-2, PhysRevB.100.165128, PhysRevB.102.195109} that allow one to construct a diagrammatic expansion based on a generic interacting reference system~\cite{BRENER2020168310}. The choice of the reference system depends on the considered lattice problem and can be, e.g., a single-site DMFT impurity problem~\cite{PhysRevB.100.205115, PhysRevB.103.245123, 10.21468/SciPostPhys.13.2.036}, several isolated impurities~\cite{PhysRevB.97.115150, 10.21468/SciPostPhys.13.2.036, vandelli2024doping}, or a cluster problem~\cite{BRENER2020168310, iskakov2024perturbative}. In this study, we adopt an effective DMFT-like impurity problem as the reference system
\begin{align}
{\cal S}_{\rm imp} = &- \int_{\cal C} dz  \,\Bigg\{  - \sum_{\sigma} c^{*}_{\sigma}(z) \left[(i\partial_{z}+\mu) \right]
c_{\sigma}^{\phantom{*}}(z) + U n_{i\uparrow}(z) n_{i\downarrow}(z) \Bigg\} \nonumber \\ 
&- \int_{\cal C} dz \, dz' \, \Bigg\{ \sum_{\sigma} c^{*}_{\sigma}(z) \, \Delta(z,z')\, c^{\phantom{*}}_{\sigma}(z')  \Bigg\}. 
\label{eq:action_imp}
\end{align}
To isolate the reference system from the initial lattice problem, we add the fermionic hybridization function $\Delta(z,z')$ to the single-particle part of the lattice action~\eqref{eq:action_latt}. 
To maintain consistency, the same hybridization function is subtracted from the remaining part of the lattice action ${\cal S}_{\rm rem} = {\cal S} - \sum_{i}{\cal S}_{\rm imp}$:
\begin{widetext}
\begin{align}
{\cal S}_{\rm rem} = - \int_{\cal C} dz \, dz' \, &\Bigg\{  - \sum_{ij,\sigma} c^{*}_{i\sigma}(z) \left[\delta_{ij}\Delta(z,z') - \delta_{\cal C}(z,z') J^{\phantom{*}}_{ij}(z)\right]
c_{j\sigma}^{\phantom{*}}(z') + \frac12 \sum_{ij,\varsigma} \rho^{\varsigma}_{i}(z) V^{\varsigma}_{ij}  \rho^{\varsigma}_{j}(z') \Bigg\}.
\label{eq:action_rem}
\end{align}
\end{widetext}
To account for the impact of the local correlations exactly, the reference problem should be integrated out.
In order to do so, 
the remaining part of the action is transformed to an effective (dual) space using Hubbard-Stratonovich transformations that introduce new fermionic ${c^{(*)} \to f^{(*)}}$ and bosonic ${\rho^{\varsigma} \to b^{\varsigma}}$ variables.
After some path integral transformation, that are discussed in detail in Refs.~\cite{PhysRevB.100.205115, PhysRevB.103.245123, 10.21468/SciPostPhys.13.2.036}, we arrives at
the effective fermion-boson (partially bosonized) action
\begin{align}
{\cal S}_{fb}
=
&\int_{\cal C} dz \, dz' \sum_{ij,\sigma} f^{*}_{i\sigma}(z) \big[\tilde{\cal G}\big]^{-1}_{ij}(z,z')
f^{\phantom{*}}_{j\sigma}(z') \nonumber \\
& + \frac12 \int_{\cal C} dz \, dz' \sum_{ij, \varsigma} b^{\varsigma}_{i}(z) \big[\tilde{\cal W}^{\varsigma}\big]^{-1}_{ij}(z,z') b^{\varsigma}_{j}(z') \nonumber \\ 
& - \int_{\cal C} dz \, dz'  dz'' \sum_{i,\varsigma} \sum_{\sigma\sigma'} \Lambda^{\varsigma}_{zz'z''} f^{*}_{i\sigma}(z) \sigma^{\varsigma}_{\sigma\sigma'} f^{\phantom{*}}_{i\sigma'}(z') b^{\varsigma}_{i}(z'').
\label{S_DB_fin}
\end{align}
with the bare dual Green's function $\tilde{\cal G}$, renormalized interaction $\tilde{\cal W}$, and the three-point vertex $\Lambda$.
The bare dual Green's function $\tilde{\cal G}$ and renormalized interaction $\tilde{\cal W}$ in the momentum-space representation explicitly read  
\begin{align}
\tilde{\cal G}_{\bf k}(z,z') &= G_{\bf k}^{\rm DMFT}(z,z') - g(z,z'),
\label{eq:bare_G} \\
\tilde{\cal W}^{\varsigma}_{q}(z,z') &=  W^{\rm EDMFT}_{q,\varsigma}(z,z') -\delta_{\cal C}(z,z') U^{\varsigma}/2.
\label{eq:bare_w}
\end{align}
Here, $g(z,z')$ is the exact impurity Green's function and ${U^{\rm d/m} = \pm U/2}$.
The DMFT Green's function $G_{\bf k}^{\rm DMFT}$ and the EDMFT-like renormalized interaction $W^{\rm EDMFT}_{q,\varsigma}(z,z')$ are given by following Dyson equation
\begin{align}
\big[G_{\bf k}^{\rm DMFT}\big]^{-1}(z,z') &= \big[{\cal G}_{\bf k}\big]^{-1}(z,z')-\Sigma^{\rm imp}(z,z'),
\label{eq:imp_G} \\
\big[W^{\rm EDMFT}_{\bf q,\varsigma}\big]^{-1}(z,z') &= \delta_{\cal C}(z,z')\big[U^{\varsigma}+V^{\varsigma}_{q}\big]^{-1} -\Pi^{\varsigma\,\rm imp}(z,z'),
\label{eq:imp_w}
\end{align}
where ${\cal G}_{\bf k}(z,z')$ is the bare lattice Green's function, and $\Sigma^{\rm imp}(z,z')$ and $\Pi^{\varsigma\,\rm imp}(z,z')$ are the self-energy and the polarization operator of the impurity problem. From the derived fermion-bosonic action~\eqref{S_DB_fin}, the self-energy and the polarization operator of the \mbox{D-TRILEX} theory in the momentum space representation can be found as follows:
\begin{widetext}
\begin{align}
\tilde{\Sigma}_{{\bf k}}(z_1,z_2)
        &= i\int_{\cal C} \{dz'\} \sum_{{\bf q},\varsigma}  \Lambda^{\varsigma}(z_1,z',z'') \, \tilde{G}^{\phantom{*}}_{{\bf k+q}}(z'',z'''') \, \tilde{W}^{\varsigma}_{{\bf q}}(z''',z'''') \, \Lambda^{\varsigma}(z'''',z_2,z''') \notag\\
&~~~-2i\int_{\cal C} \{dz'\} \sum_{{\bf k'}} \Lambda^{\rm d}(z_1,z_2,z'') \, \tilde{\cal W}^{\rm d}_{{\bf q}=0}(z'',z''') \, \Lambda^{\rm d}(z''',z'''',z'') \, \tilde{G}^{\phantom{*}}_{{\bf k'}}(z'''',z''') \,,
\label{eq:Sigma_dual_actual}\\
\tilde{\Pi}^{\varsigma}_{{\bf q}}(z_1,z_2)
&= -2i \int_{\cal C} \{dz'\} \sum_{{\bf k}} \Lambda^{\varsigma}(z'',z',z_1) \, \tilde{G}^{\phantom{*}}_{{\bf k}}(z',z''') \, \tilde{G}^{\phantom{*}}_{{\bf k+q}}(z'''',z'') \, \Lambda^{\varsigma}(z''',z'''',z_2) \,,
\label{eq:Pi_dual_actual}
\end{align}
\end{widetext}
where the dressed dual Green's function and renormalized interactions on Keldysh contour are defined as 
\begin{align}
\tilde{G}_{\bf k}(z,z') &= -i\langle T_{\cal C} f^{\phantom{*}}_{\bf k,\sigma}(z) f^{*}_{\bf k,\sigma}(z')\rangle, \\
\tilde{W}^{\varsigma}_{\bf q}(z,z') &= -i\langle T_{\cal C} b^{\varsigma}_{\bf{q}}(z)b^{\varsigma}_{\bf{q}}(z')\rangle.
\end{align}
The diagrammatic structure of dual self-energy and dual polarization of D-TRILEX is given by 
\begin{equation}
  \tikz[baseline, anchor=base]{ \begin{scope}[yshift=0.08cm]
    \draw (0, 0) node[circle, draw, fill=Red!50, anchor=center] {$\tilde{\Sigma}$};
    \draw (0.6, -0.08cm) node[circle] {$=$};
    \draw (1.2,0) node[name=v2, TriangleVertex, shape border rotate=60] {};
    \draw (1.2,0.75) node[name=v1, TriangleVertex, shape border rotate=60+180] {};
    \draw [interaction] (v1.apex) -- (v2.apex);
    \draw [electronBold] (v1.left corner) to[distance=0.5cm, out=+60, in=+120] (v1.right corner);
    \draw (1.7, -0.08cm) node[] {$+$};
    \draw (2.2,0) node[name=v3, TriangleVertex, shape border rotate=60] {};
    \draw (3.5,0) node[name=v4, TriangleVertex, shape border rotate=60] {};
    \draw [interactionBold] (v3.apex) to[distance=0.75cm, out=90, in=90] node[midway, below=2mm] {\scriptsize $\text{ch+sp}$}(v4.apex);
    \draw [electronBold] (v4.left corner) -- (v3.right corner);
  \end{scope}}
  \, , \quad
  \tikz[baseline, anchor=base]{ \begin{scope}[yshift=0.08cm]
    \draw (0, 0) node[draw, minimum size=0.65cm, rounded corners, fill=Blue!20, name=S, anchor=center] {$\tilde{\Pi}^\varsigma$};
    \draw (0.6, -0.08cm) node[] {$=$};
    \draw (0.9,0) node[name=v2, TriangleVertex, shape border rotate=180] {};
    \draw (2.5,0) node[name=v1, TriangleVertex, shape border rotate=0] {};
    \draw [electronBold] (v1.left corner) -- (v2.right corner);
    \draw [electronBold] (v2.left corner) -- (v1.right corner);
  \end{scope}}
  \, ,
\end{equation}
where
$
\Lambda = 
\tikz[baseline, anchor=base]{ \begin{scope}[yshift=0.09cm]
\draw (0,0) node[name=v, SmallTriangleVertex, rotate=0] {};
\end{scope}}
$ 
is the three-point vertex,
$
\tilde{W}^\varsigma_{\mathbf{q}} = \,
  \tikz[baseline, anchor=base]{ \begin{scope}[yshift=0.08cm]
    \draw [interactionBold] (1.5,0) -- node[midway, above] {\scriptsize $\varsigma$} (0,0);
  \end{scope}}
  \, 
$
and 
$
\tilde{G}_{\mathbf{k},\sigma} = \,
  \tikz[baseline, anchor=base]{ \begin{scope}[yshift=0.08cm]
    \draw [electronBold] (1.5,0) -- (0,0);
  \end{scope}}
  \, 
$
are the dressed dual renormalized interaction and Green's function, respectively.
$
\mathcal{\tilde{W}}^{\varsigma}_{\mathbf{q}} = \,
  \tikz[baseline, anchor=base]{ \begin{scope}[yshift=0.08cm]
  \draw [interaction] (1.5,0) -- node[midway, above] {\scriptsize $\varsigma$} (0,0);
  \end{scope}}
  \, ,
$ is the bare dual renormalized interaction. 

\begin{figure*}[t!]
\includegraphics[width=0.92\linewidth]{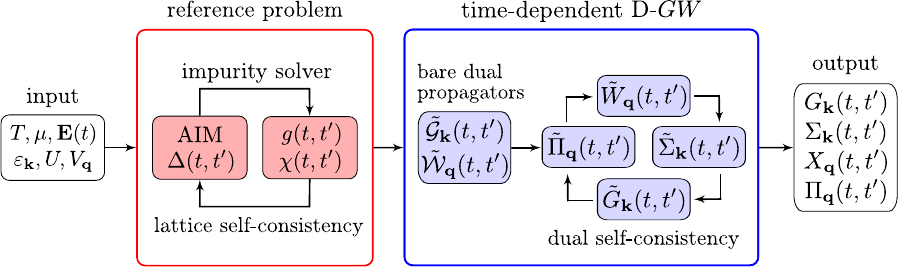}
\caption{The computational workflow of the non-equilibrium \mbox{D-$GW$} method on Kelydsh contour $z=t \in C_{\pm,M}, z'=t' \in C_{\pm, M}$. 
The input consists of the model parameters: the initial temperature $T$, the chemical potential $\mu$, the applied electric field ${\bf{E}}_{p}(t)$, the electronic dispersion $\varepsilon_{\bf k}$, and the local $U$ and non-local $V_{\bf q}$ electronic interactions. The red box indicates the self-consistent solution of the time-dependent reference problem. This work considers the latter an Anderson impurity model (AIM) with a time-dependent hybridization function $\Delta(t,t')$ determined by the DMFT lattice self-consistency. The \mbox{D-$GW$} formalism (blue box) takes the renormalized Green's function $g(t,t')$ and susceptibilities $\chi(t,t')$ of the reference problem are further used as inputs for constructing bare propagators for the dual fermionic $\tilde{\cal G}_{\bf k}(t,t')$ and bosonic $\tilde{\cal W}_{\bf q}(t,t')$ fields. The dressed time-dependent quantities in dual space: the polarization operator $\tilde{\Pi}_{\bf q}(t,t')$, the renormalized interaction $\tilde{W}_{\bf q}(t,t')$, the self-energy $\tilde{\Sigma}_{\bf k}(t,t')$, and the renormalized Green's function $\tilde{G}_{\bf k}(t,t')$ are determined through the \mbox{D-$GW$} dual self-consistency (closed loop in the blue box). The physical quantities are obtained using the exact relations between the corresponding quantities in the dual and lattice spaces.
\label{fig:cycle}}
\end{figure*}

We note that the vertex function $\Lambda^{\varsigma}$ can be defined in several ways, and its form depends on how the Hubbard-Stratonovich transformation is performed to bosonize collective electronic fluctuations (for different definitions of the vertex see, e.g., Refs.~\cite{PhysRevB.90.235135, ayral:tel-01247625, PhysRevB.94.205110, PhysRevB.100.205115, PhysRevB.103.245123, 10.21468/SciPostPhys.13.2.036}).
These different formulations of the theory modify only the quantities in the effective (dual) space but yield identical results for the physical observables.
However, as demonstrated in Ref.~\cite{PhysRevB.100.205115}, there is a unique definition such that the asymptotic short-time (infinite frequency) limit of the three-point vertex is equal to unity. In this work, we adhere to this specific formulation of the method, which, upon approximating the vertex by its instantaneous component ${\Lambda^{\varsigma}_{z_1z_2z_3} = \delta_{\cal{C}}(z_1,z_2) \delta_{\cal{C}}(z_2,z_3)}$, yields a simple $GW$-like structure for the \mbox{D-TRILEX} self-energy and polarization operator that we refer to as \mbox{D-$GW$}: 
\begin{align}
\tilde{\Sigma}_{{\bf k}}(z_1,z_2)
        &= i \sum_{{\bf q},\varsigma} \tilde{G}^{\phantom{*}}_{{\bf k+q}}(z_1,z_2) \tilde{W}^{\varsigma}_{{\bf q}}(z_1,z_2) \nonumber \\
        -2i&\delta_{\cal C}(z_1,z_2)\sum_{\bf{k'}} \int_{\cal{C}} dz'\tilde{\cal W}^{d}_{{\bf q}=0}(z_1,z') \tilde{G}^{\phantom{*}}_{{\bf k'}}(z',z'),
\label{eq:Sigma_dual_para}\\
\tilde{\Pi}^{\varsigma}_{\bf q}(z_1,z_2)
        &= -2i \sum_{\bf {k}} \tilde{G}_{\bf {k}}(z_1,z_2) \tilde{G}_{\bf {k+q}}(z_2,z_1).
\label{eq:Pi_dual_para}
\end{align}
The dressed dual Green's function and renormalized interactions can be found by solving the following Dyson equations
\begin{align}
\big[\tilde{G}_{\bf k}\big]^{-1}(z,z') = \big[\tilde{\cal G}_{\bf{k}}\big]^{-1}(z,z') -\tilde{\Sigma}_{\bf k}(z,z'),
\label{eq:dress_G} \\
\big[\tilde{W}^{\varsigma}_{\bf q}\big]^{-1}(z,z') = \big[\tilde{\cal W}^{\varsigma}_{\bf q}\big]^{-1}(z,z')-\tilde{\Pi}^{\varsigma}_{\bf q}(z,z').
\label{eq:dress_w}
\end{align}

The \mbox{D-$GW$} theory has several benefits since the diagrammatic expansion is performed in the dual space. First, independently of the diagrammatic approximation, the theory is free from the double-counting of the correlation effects between the reference (impurity) and remaining (diagrammatic) parts of the problem. 
Second, such an expansion becomes exact in both the weak and strong coupling limits, simultaneously combining essentially different perturbation expansions used in these two limits. 
Finally, dual diagrams allow one to avoid another double-counting issue, known as Fierz ambuguity~\cite{PhysRevD.68.025020, PhysRevB.70.125111, Jaeckel}, that appears when collective electronic fluctuations in several interplaying (e.g., charge and spin) channels are considered simultaneously~\cite{PhysRevLett.119.166401}.

The computational workflow of \mbox{D-$GW$} is illustrated in Fig.~\ref{fig:cycle}. 
The first step (red box) involves the self-consistent solution of the reference (DMFT impurity) problem, which, in this work, is done using a lowest-order strong coupling perturbative method called the real-time non-crossing approximation (NCA). The formulation of real-time NCA on the Keldysh contour is detailed in Ref.~\cite{Mott-NCA, Hugo2017PRB}, while the real-time DMFT self-consistency implementation on a square lattice is described in Appendix~\ref{app:DMFT}. Note, that the impurity problem does not contain the effect of the non-local interaction ${(V=0)}$.
The solution of the reference problem, subject to an external time-dependent perturbation ${{\bf{E}}_{p}(t)}$, yields the two-time dependent Green's function $g(t,t')$ and the susceptibilities $\chi^{\varsigma}(t,t')$. 
Together with the self-consistently obtained hybridization function $\Delta(t,t')$, these quantities are further used in the \mbox{D-$GW$} diagrammatic expansion to construct the bare dual Green's function $\tilde{\cal G}_{\bf k}(t,t')$~\eqref{eq:bare_G} and renormalized interaction $\tilde{\cal W}^{\varsigma}_{\bf q}(t,t')$~\eqref{eq:bare_w}.
The dressed dual quantities are obtained via self-consistent solution of the corresponding Dyson equations~\eqref{eq:dress_G} and~\eqref{eq:dress_w} that involve the diagrammatic calculation of the dual self-energy $\tilde{\Sigma}_{\bf k}(t,t')$~\eqref{eq:Sigma_dual_para} and polarization operator $\tilde{\Pi}^{\varsigma}_{\bf q}(t,t')$~\eqref{eq:Pi_dual_para}. 
Ultimately, the dressed physical Green's function $G_{\bf k}(t,t')$ and susceptibilities $X^{\varsigma}_{\bf q}(t,t')$, as well as the self-energy $\Sigma_{\bf k}(t,t')$ and polarization operator $\Pi_{\bf q}(t,t')$ are obtained using the exact relations between the dual and lattice quantities~\cite{PhysRevB.100.205115, PhysRevB.103.245123, 10.21468/SciPostPhys.13.2.036}.
The latter are explicitly defined in the next section.

\subsection{Physical Green's functions in \mbox{D-$GW$}}

In the following sections, we will present the 
quantities of interest for data analysis. 
The first and most crucial quantity of interest is the lattice Green's function 
\begin{align}
G_{\bf k}(z,z') = -i\langle T_{\cal C} c^{\phantom{\dagger}}_{\bf k,\sigma}(z) c^{\dagger}_{\bf k,\sigma}(z')\rangle,  
\label{eq:G_latt_def}
\end{align}
which is given by the following integral form: 
\begin{widetext}
\begin{align}
\int_{\cal C} dz'' \, dz''' \left\{\delta_{\cal C}(z,z''') + \big[g(z,z'') + \tilde{T}_{\bf k}(z,z'')\big]\big[\Delta(z'',z''') - (\epsilon_{\bf k}(z)-\epsilon_{\rm loc}(z))\delta_{\cal C}(z'',z''')\big] \right\} G_{\bf k}(z''',z') = g(z,z') + \tilde{T}_{\bf k}(z,z'),
\label{eq:G_latt}
\end{align}
\end{widetext}
where ${\tilde{T}_{\bf k}(z,z') = \int_{\cal C} dz'' \, dz''' \, g(z,z'') \tilde{\Sigma}_{\bf k}(z'',z''')g(z''',z')}$. 

The lattice susceptibility on Keldysh contour is defined as
\begin{align}
X^{\varsigma}_{\bf q}(z, z') =
-i\left\langle
T_{\cal C} \,
\rho^\varsigma_{\bf q}(z) 
\,
\rho^\varsigma_{\bf -q}(z') 
\right\rangle.
\end{align} 
To calculate this quantity, one has to obtain the lattice polarization operator
\begin{align}
\Pi^{\varsigma}_{\bf q}(z,z') = \Pi^{\varsigma}_{\rm imp}(z,z') + \overline{\Pi}^{\varsigma}_{\bf q}(z,z'),
\end{align}
where the diagrammatic correction $\overline{\Pi}^{\varsigma}_{\bf q}(z,z')$ to the impurity polarization operator $\Pi^{\varsigma}_{\rm imp}(z,z')$ reads:
\begin{align}
\overline{\Pi}^{\varsigma}_{\bf q}(z,z') = \tilde{\Pi}^{\varsigma}_{\bf q}(z,z') - \int_{\cal C} dz'' \, \tilde{\Pi}^{\varsigma}_{\bf q}(z,z'') \frac{U^{\varsigma}}{2} \overline{\Pi}^{\varsigma}_{\bf q}(z'',z').
\end{align}
The lattice susceptibility can then be obtained via the Dyson equation
\begin{align}
X^{\varsigma}_{\bf q}(z,z') = \Pi^{\varsigma}_{\bf q}(z,z') + \int_{\cal C} dz'' \, \Pi^{\varsigma}_{\bf q}(z,z'') \left(U^{\varsigma} + V^{\varsigma}_{\bf q}\right) X^{\varsigma}_{\bf q}(z'',z').
\label{eq:chi_latt}
\end{align}

The lattice self-energy:
\begin{align}
\Sigma^{\rm latt}_{{\bf k}}(z,z') = \Sigma^{\rm imp}(z,z') + \overline{\Sigma}_{{\bf k}}(z,z').
\end{align}
consists of the impurity contribution $\Sigma^{\rm imp}(z,z')$ and the diagrammatic correction $\overline{\Sigma}_{{\bf k}}(z,z')$.
The latter can be obtained as follows:
\begin{align*}
\overline{\Sigma}_{{\bf k}}(z,z') = \tilde{\Sigma}_{{\bf k}}(z,z') - \int_{\cal C} dz'' \, dz''' \, \tilde{\Sigma}_{{\bf k}}(z,z'') g(z'',z''') \overline{\Sigma}_{{\bf k}}(z''',z').
\end{align*}
Alternatively, the lattice Green's function, defined in Eq.~\ref{eq:G_latt_def}, can be calculated from the diagrammatic correction to the impurity self-energy using the following Dyson equation
\begin{align}
\big[G^{\phantom{1}}_{\bf{k}}\big]^{-1}(z,z') = \big[G^{\rm DMFT}_{\bf k}\big]^{-1}(z,z')- \overline{\Sigma}_{{\bf k}}(z,z').
\end{align}
In this work, the lattice Green's function is calculated using Eq.~\ref{eq:G_latt}, which is more efficient computationally.
\subsection{Equal time observables:}
The properties of the initial thermal equilibrium state is obtained by restricting the contour times $z, z'$ to the imaginary time contour $z = -i\tau, z' = -i\tau' \in \mathcal{C}_M$, $G_\mathbf{k}(\tau-\tau') = G_\mathbf{k}(z, z')$, while the non-equilibrium properties are obtained when restricting the contour times $z$ and $z'$ to the real-time branches,
giving the lesser and greater response function components,
\begin{equation}
G^{\lessgtr}_\mathbf{k}(t, t') =
G_{\mathbf{k}}(z, z')
\, , \quad
z = t \in \mathcal{C}_\pm
\, , \,\,
z' = t' \in \mathcal{C}_\mp
\,.
\end{equation}
which corresponds to the occupied and unoccupied states. The equal-time values of the propagators give physical observables, e.g.\ the time-dependent electron density ${\langle n_{\mathbf{k}\sigma}(t) \rangle = -iG^<_{\mathbf{k}}(t, t^+)}$, that determines the kinetic energy
\begin{equation}
K(t) = \frac{2}{N_\mathbf{k}} \sum_{\mathbf{k}} \varepsilon_{\mathbf{k}}(t) \langle n_{\mathbf{k}\sigma} (t) \rangle
\, .
\label{eq:Ekin}
\end{equation}
The interaction energy within the Galitskii-Migdal formula is given by the convolution~\cite{Mahan}
\begin{equation}
P(t) = 
-\frac{i}{N_\mathbf{k}} \sum_{\mathbf{k}} [\Sigma_{\mathbf{k}} \ast G_{\mathbf{k}}]^<(t, t^+),
\label{eq:Eint}
\end{equation}
where $\Sigma_{\mathbf{k}}(z, z')$ is the single-particle self-energy and the contour product is given by $(A \ast B)(z, z') \equiv \int_\mathcal{C} d\bar{z} A(z, \bar{z})B(\bar{z}, z')$, see~\cite{Aoki2014}.
The combination gives the total energy 
\begin{equation}
E(t) = K(t) + P(t)
\, .
\label{eq:Etot}
\end{equation}
The interaction energy of the introduced model~\eqref{eq:model} can be separated into local $E_{U}$ and non-local $E_{V}$ contributions, where the former is due to the Hubbard interaction $U$, and the latter comes from the nearest-neighbor non-local interaction $V$:
\begin{equation}
P(t) = E_{U}(t) + E_{V}(t).
\label{eq:total_E}
\end{equation}
The local interaction energy is, in turn, given by ${E_U(t) = U D(t)}$, where the local double occupancy $D(t)$ can be determined using the local charge and spin susceptibilities $\chi^\varsigma$~\cite{Erik2016},
\begin{align}
D(t) &\equiv 
\langle n_{i\uparrow}(t) n_{i\downarrow}(t) \rangle \notag\\
&= 
\frac{1}{4}\left(
iX^{d,<}_{\rm loc}(t, t^+) 
- 
iX^{m,<}_{\rm loc}(t, t^+)
+
{\langle n^{d}(t) \rangle}^{2} 
\right).
\end{align}
The above double occupancy expression can be further used to decouple the local interaction energy between charge and spin channels ${E_{U} = E_{Ud} + E_{Um}}$, where 
\begin{align}
E_{Ud} &= \frac{U}{4} \left( iX^{d,<}_{\rm loc}(t, t^+) + \langle n^{d}(t) \rangle^2 \right), \\ 
E_{Um} &= -\frac{U}{4} iX^{m,<}_{\rm loc}(t, t^+).  
\end{align}
Finally, the non-local interaction energy can be obtained from the momentum-dependent charge susceptibility as follows:
\begin{equation}
E_{V}(t)= \frac12 \sum_{\bf q} V_{\bf q} iX^{d,<}_{\bf q}(t,t^+).
\end{equation}

Since the interaction energy can be calculated using either the self-energy (via the Galitskii-Migdal formula) or the susceptibilities, the total energy of the photo-excited state can be evaluated in two distinct ways.
In this paper, we denote the total energy as \(E_{\chi}\) when the interaction energy is calculated from the  susceptibility and as \(E_{Mig}\) when calculated from the self-energy. 
The non-equilibrium Green's function formalism allows us to access all components of the system's total energy. This capability enables us to study the non-equilibrium energy redistribution during and after photo-doping in the system.

\subsection{Spectral functions}
The non-equilibrium Green's function formalism provides access to real-time spectral information of electrons, as well as charge and spin degrees of freedom. The real-time generalization of the occupied and unoccupied electronic spectral functions is obtained using a partial Fourier transform
\begin{equation}
A_\mathbf{k}^{\lessgtr}(t, \omega) = \frac{\pm\text{Im} \hspace{0.05cm}G^{\lessgtr}_{\mathbf{k}}(t,\omega)}{\pi} = 
\int_0^\infty \!\!\!\! d\bar{t} \, 
e^{i\omega \bar{t}} \, G^{\lessgtr}_\mathbf{k}(t + \bar{t}, t) 
\, ,
\end{equation}
which combines into the time-dependent electronic spectral function
\begin{equation}
A_{\mathbf{k}}(t, \omega) = 
A^{>}_{\mathbf{k}}(t, \omega) + A^{<}_{\mathbf{k}}(t, \omega)
\, .
\end{equation}
At thermal equilibrium, the time dependence dropout and the fluctuation-dissipation theorem gives
\begin{equation}
A_\mathbf{k}^<(\omega) = f_{e}(\omega) A_\mathbf{k}(\omega)
\, , ~\,
A_\mathbf{k}^>(\omega) = \left[ 1 - f_{e}(\omega) \right] A_\mathbf{k}(\omega)
\, ,
\end{equation}
where $f_{e}(\omega) = 1/(e^{\beta\omega} - 1)$ is the Fermi distribution function. 
This results in $F_{k}(\omega) \equiv \ln \left[ A^{<}_{k}(\omega) / A^{>}_{k}(\omega) \right] = -\omega \beta$.

As used for electrons, a similar Fourier transform has been used to calculate the occupied and unoccupied spectral functions of charge and spin-susceptibility. They are defined as
\begin{equation}
A^{d/m,\lessgtr}_\mathbf{q}(t, \omega) = \frac{-\text{Im} \hspace{0.05cm}X^{d/m,\lessgtr}_{\mathbf{q}}(t,\omega)}{\pi} = 
\int_0^\infty \!\!\!\! d\bar{t} \, 
e^{i\omega \bar{t}} \, X^{d/m,\lessgtr}_\mathbf{q}(t + \bar{t}, t) 
\, .
\end{equation}
One can calculate the retarded spectral functions of charge and spin susceptibilities from occupied and un-occupied functions, which are given by 
\begin{equation}
A^{d/m}_{\mathbf{q}}(t, \omega) = 
A^{d/m,>}_{\mathbf{q}}(t, \omega) - A^{d/m,<}_{\mathbf{q}}(t, \omega)
\, .
\end{equation}
The fluctuation-dissipation theorem in thermal equilibrium relates the retarded spectral functions to their occupied and unoccupied susceptibilities through
\begin{equation}
A^{d/m,<}_\mathbf{q}(\omega) = f_{b}(\omega) A^{d/m}_\mathbf{q}(\omega)
\, , ~\,
A^{d/m,>}_\mathbf{q}(\omega) = \left[ 1 - f_{b}(\omega) \right] A^{d/m}_\mathbf{q}(\omega)
\, ,
\end{equation}
where ${f_{b}(\omega) = 1/(e^{\beta\omega} + 1)}$ is the Bose distribution function. 
The occupied-to-unoccupied spectral susceptibilities ratio leads to a similar electron expression. To explore this out-of-equilibrium, we use the time-dependent generalization 
\begin{align}
F_{k}(t, \omega) \equiv \ln \left[ \frac{A^{<}_{\mathbf{k}}(t, \omega)}{A^{>}_{\mathbf{k}}(t, \omega)} \right],
~~F^{d/m}_{q}(t,\omega) =\ln \left[ \frac{A^{d/m,<}_{\mathbf{q}}(t, \omega)}{A^{d/m,>}_{\mathbf{q}}(t, \omega)} \right],
\label{hkjwdlnsks}
\end{align}
which allows us to study the system's approach to thermalization. In the long-time limit, as ${t \rightarrow \infty}$, we find that $F_{k}(t, \omega)$ and $F^{d/m}_{q}(t,\omega)$ approach to  -$\beta_{\text{eff}}(t) \omega$, enabling us to determine the final effective inverse temperature $\beta_{\text{eff}}$. Additionally, it has been utilized to extract effective temperatures for quasi-particles, doublons, and charge and spin degrees of freedom during the relaxation of photo-excited states \cite{Dasari2018, Sayyad, Non-eq_book}.

\subsection{Numerical setup}

We use the NESSi implementation~\cite{SCHULER2020107484} for the solution of the Kadanoff-Baym equations on the Keldysh contour. Unless otherwise stated, we fix the initial equilibrium temperature of the systems at ${T=1/\beta=1/6}$ and pulse frequency to ${\omega_{p}=U}$. Calculations are performed for the periodic lattice of size ${20\times20}$. We further increase the lattice size to ${26\times26}$ and confirm that our results are independent of the system size. We select the Matsubara grid with $\Delta \tau$ = 0.0066 (0.0060) and the real-time grid with $\Delta t$ = 0.011 (0.010) for metals (Mott-insulators) to ensure that dynamical quantities converge with grid size.

\section{Equilibrium \label{sec:equilibrium}}

Before discussing the photo-excitation dynamics of the extended Hubbard model~\eqref{eq:model}, let us first focus on the equilibrium results to determine the effect of non-local correlations, particularly spin fluctuations, on the electronic spectral function. 
These results will also provide us with further insights to identify the parameter space of interest for the photo-excitation dynamics.
Without the electric pulse ${\mathbf{E}_{p}(t)=0}$, the solution of the model on the $L$-shaped contour yields a time-translationally invariant electronic Green's function~\eqref{eq:G_latt} and susceptibilities~\eqref{eq:chi_latt}, which are the functions of time-difference ${(t-t')}$.
A straightforward Fourier transformation of these functions from real-time to frequency domain directly provides access to spectral information without requiring analytical continuation in finite-temperature Matsubara methods.

\begin{figure}[t!]
\includegraphics[width=0.9\linewidth, trim={10pt 10pt 10pt 10pt}, clip]{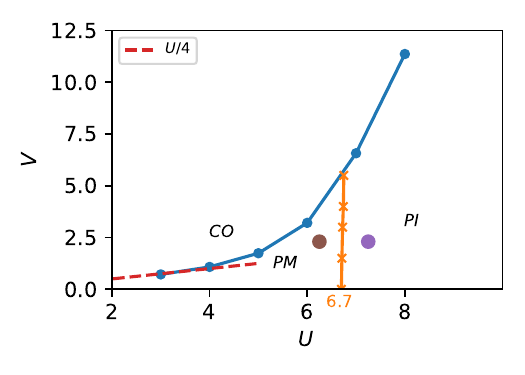}
\caption{Equilibrium phase diagram of the extended Hubbard model in the ${U\text{-}V}$ plane obtained using the real-time \mbox{D-$GW$} method.
The blue line indicates the phase boundary between the charge-ordered (CO) and paramagnetic metal (PM) phases. 
The orange line depicts the phase boundary between the PM and the paramagnetic Mott insulator (PI). 
The red dashed line depicts the mean-field estimate ${V=U/4}$ for the CO phase boundary.
The brown (${U=6.25}$, ${V=2.3}$) and purple circles (${U=7.25}$, ${V=2.3}$) respectively indicate the parameters of a correlated metal and a Mott insulator at which calculations are carried out in this paper.
\label{fig:fig1}}
\end{figure}

In Fig.~\ref{fig:fig1} we show the phase diagram of the extended Hubbard model at half-filling in the ${U\text{-}V}$ plane  
obtained using the developed real-time \mbox{D-$GW$} method. 
Without the non-local interaction (${V=0}$), the problem reduces to the single-band Hubbard model, which in the half-filled case has been extensively studied using various numerical methods. 
At a critical temperature $T_N$, this model features the N\'eel transition to the antiferromagnetic (AFM) state driven by strong spin fluctuations.
Above $T_N$, the long-range AFM order completely melts, but the short-range spin fluctuations persist to higher temperatures in the paramagnetic phase. Since the equilibrium temperature is chosen to be larger than AFM N\'eel temperature ($T>T_{N}$), we can not see such an AFM to paramagnetic transition in Fig.~\ref{fig:fig1}.
At a critical interaction $U_{c}$, the high-temperature paramagnetic phase exhibits the transition to the Mott insulating state, which is associated with the effect of local electronic correlations.
The formation of the insulating state driven by the magnetic fluctuations (Slater mechanism) and local electronic correlation (Mott scenario) can be distinguished by how the gap forms in the electronic spectral function, which is momentum-selective in the former case and momentum-independent in the latter case~\cite{PhysRevLett.132.236504}.

Following Ref.~\onlinecite{PhysRevLett.132.236504}, we characterize the Mott transition through the simultaneous disappearance of the low-energy coherent peak in the electronic spectral function $A_{\bf k}(\omega)$
at the nodal (${\text{N}=(\pi/2,\pi/2)}$) and anti-nodal (${\text{AN}=(0,\pi)}$) points of the Fermi surface. 
At ${V=0}$, the real-time \mbox{D-$GW$} study of the Hubbard model 
predicts Mott-transition at ${U_{c}=6.7}$. Usually, considering the non-local interaction $V$ results in an effective screening of the on-site potential $U$, which leads to an increase of the critical interaction $U_{c}$ for the Mott transition with increasing $V$ (see, e.g, Refs.~\onlinecite{PhysRevB.66.085120, PhysRevB.87.125149, PhysRevB.90.195114}). 
However, Fig~\ref{fig:fig1} shows that the obtained phase boundary of the Mott transition is nearly independent of the value of $V$. We attribute this result to the lack of vertex corrections neglected in the simplified version of the \mbox{D-$GW$} method. In addition to screening $U$, the nearest-neighbor interaction $V$ 
on a square lattice favors the formation of a charge-ordered (CO) phase.
At half filling, this state is characterized by a checkerboard pattern of alternating empty and doubly occupied lattice sites and has also been studied using a variety of methods~\cite{PhysRevB.66.085120, PhysRevB.87.125149, PhysRevB.90.195114, PhysRevB.90.235135, PhysRevB.94.205110, PhysRevB.95.245130, PhysRevB.95.115149, PhysRevB.97.115117, PhysRevB.99.245146, PhysRevB.100.165128, PhysRevB.102.195109}.
The phase transition to the CO state can be detected by the divergence of the static charge susceptibility ${X^{d}_{{\bf q}}}$ at the ${{\bf q} = (\pi,\pi)}$ point. 
At small values of $U$ the phase boundary of the CO phase predicted by \mbox{D-$GW$} matches with the mean-field result ${V = U/4}$ (dashed line).
At larger interactions, the phase boundary is shifted above this line, in agreement with previous studies.

\begin{figure}[t!]
\includegraphics[width=0.9\linewidth, trim={10pt 10pt 10pt 10pt}, clip]{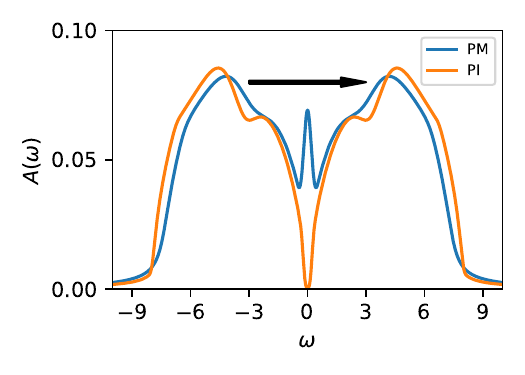}
\caption{Equilibrium electronic spectral functions of paramagnetic metals (PM) and paramagnetic Mott insulators (PI).\label{fig:optics}}
\end{figure}

Let us focus on one of the most challenging regimes for theoretical analysis, namely the region  near the Mott and CO phase transitions. In this regime, local electronic correlations exhibit a non-perturbative character. 
Additionally, the system displays strong non-local charge and spin fluctuations on the metallic side.
To determine the effect of spatial collective electronic fluctuations, we calculate the local electronic spectral function ${A(\omega) = \frac{1}{N_k}\sum_{\bf k}A_{\bf k}(\omega)}$ for the two points depicted in Fig~\ref{fig:fig1} by the circles.  
The brown circle represents the correlated metal (${U=6.25}$, ${V=2.3}$), and the purple one corresponds to the narrow-gap Mott insulator (${U=7.25}$,${V=2.3}$); see Fig.~\ref{fig:optics} for the corresponding equilibrium spectral functions. As expected, a three-peak structure with lower and upper Hubbard bands, along with a quasi-particle peak near the Fermi level—known as the Abrikosov-Shul\' resonance\cite{Abrikosov1965} in correlated metals—distinguishes them from Mott insulators, where a narrow gap near the Fermi level is identified. As we further focus on the effect of spin fluctuations, the interaction parameters are chosen to make short-range spin fluctuations more significant than the charge ones.

\section{Photo doping\label{sec:photo-doping}}

\subsection{Pump protocol}

\begin{figure}[t!]
\includegraphics[width=0.95\linewidth]{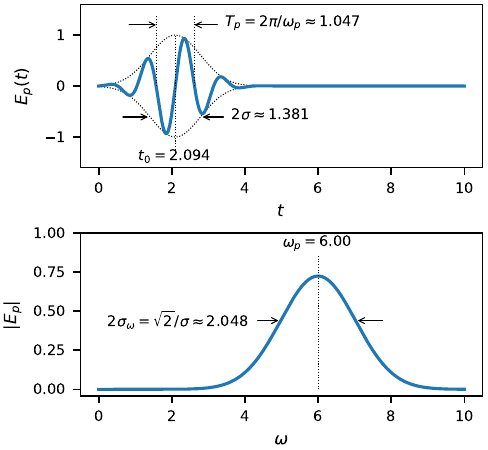} 
\caption{Time dependent electrical field $E_{p}(t)$ of the applied laser pump in time (upper panel) and its frequency power spectra (lower panel). 
\label{fig:pulse_new}}
\end{figure}

The electrical field of the applied laser pulse is directed along the lattice diagonal and has the form
\begin{equation}
    \mathbf{E}_{p}(t) = 
    \hat{n}_{xy} \cdot 
    E_0 \sin( \omega_p [t - t_0] ) \cdot e^{-\frac{(t - t_0)^2}{2\sigma^2}}
\end{equation}
with amplitude $E_0 = 1$, frequency $\omega_p = 6$, a Gaussian amplitude modulation centered at time $t_0 = 2.094$ with standard deviation $\sigma = 0.690$, and the unit vector $\hat{n}_{xy}$ in the positive x and y diagonal direction. This gives a period $T_p = 2\pi / \omega_p \approx 1.047$ and standard deviation $\sigma_\omega = 1/\sqrt{2 \sigma^2} \approx 1.024$ in frequency, see Fig.~\ref{fig:pulse_new}. Tuning the frequency \(\omega_p\) to align with the local Hubbard interaction \(U\) (i.e., ${\omega_p \approx U}$) centers the power spectra of the pump on the resonant excitation of electrons from the lower to the upper Hubbard band, as shown by the horizontal arrow in Fig.~\ref{fig:optics}. This alignment ensures that the energy quanta absorbed by the electrons, which are accelerated by the electric field, are sufficient to create local doubly occupied sites (doublons) and local empty sites (holons). In addition to the incoherent excitations, the electric pulse also influences the low-energy quasiparticle excitations commonly found in metals. These low-energy excitations can be observed in the single-particle spectral functions, as illustrated in Fig.~\ref{fig:optics}. 

We set the pulse amplitude ${E_0 = 1}$, which induces a density of 1.5\% for holons and doublons in the photo-excited state for the chosen metallic and Mott-insulating initial states through a non-linear process. The doublon density is defined by change of its value from initial equilibrium value $\Delta D(t)= \frac{D(t=4)-D(t=0)}{D(t=0)}$. There are two primary excitation mechanisms responsible for producing doublon-holon excitations under an ac-field drive, both governed by the Keldysh parameter ${\gamma \equiv \frac{\omega_{p}}{E_0 \xi}}$, where $\xi$ characterizes the spatial correlation length of doublon-holon pair~\cite{Keldysh}. For the parameters of ${E_0}$ and ${\omega_{p}}$ selected in this study, the multi-photon absorption process (where ${\gamma \gg 1}$) is more dominant than quantum tunneling (where ${\gamma \ll 1}$). In the former case, the production rate of doublons demonstrates a power-law dependence on the electric field strength $E_0$. Additionally, the density of photoexcited doublons and holons in the multi-absorption process exhibits a momentum-dependent distribution, with doublons and holes primarily created at the edges of the Mott gap~\cite{Oka2012}.

\subsection{ Energy conservation in the photo-excited state}

We analyze many-body theory in the context of an electric pulse by examining conservation laws. Among these, the most significant is total energy, which is crucial for understanding the relaxation dynamics of photo-excited charge carriers in a closed quantum system. The relationship between total energy $E(t)$ and an applied electric field $\mathbf{E}_{p}(t)$ is described by the equation ${\frac{dE(t)}{dt} = j(t) \cdot \mathbf{E}_{p}(t)}$, where $j(t)$ is the photo-induced electric current in the system. In the absence of an electric pulse, the total energy remains constant over time, as indicated by ${\frac{dE(t)}{dt} = 0}$. 

\begin{figure}[t!]
\includegraphics[width=0.88\linewidth, trim={10pt 10pt 10pt 10pt}, clip]{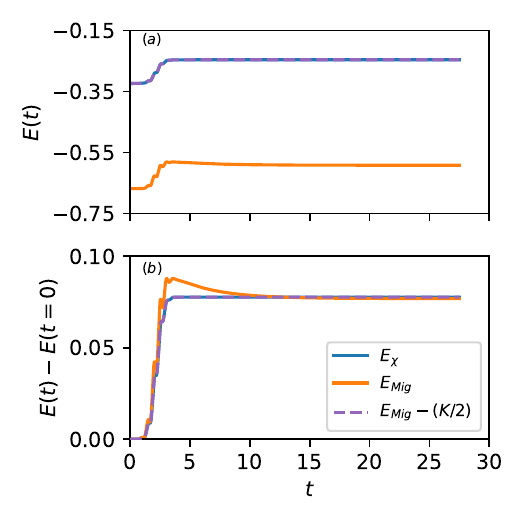}
\caption{(a) The total energy of the photo-doped metals within DMFT is plotted in the upper panel, and their change from the initial equilibrium state is plotted in the lower panel. 
\label{fig:total_energies_dmft}}
\end{figure}

DMFT is known to be a conserving approximation, as described by Baym and Kadanoff, where the many-body self-energy is connected to the Green's function through the derivative of the Luttinger-Ward function~\cite{Baym-Kadanoff}.
We examined the conservation of the total energy in DMFT, as illustrated in Fig.~\ref{fig:total_energies_dmft}(a). 
As anticipated, the total energy calculated from the two-particle susceptibility, denoted as ${E_{\chi}}$, reached a constant value immediately after the pulse (for ${t < 4}$), confirming energy conservation. 
The total energy calculated from the electronic self-energy, labeled \(E_{Mig}\), also appears to be conserved after the pulse. However, there is a notable difference in the magnitude of the total energy derived from these two approaches. 
In Fig.~\ref{fig:total_energies_dmft}(b), we further plot the change in the total energy from the initial value  for clear identification of dynamics that occurs immediately after the pulse. 
We find that the quantity \(E_{Mig}\) violates energy conservation right after the pulse but maintains conservation for extended periods, while the ${E_{\chi}}$ becomes constant immediately when the pulse is switched of. 
DMFT is a conserving theory by construction, and both quantities should therefore be conserved and yield similar magnitudes.
However, our numerical calculations are based on the approximate (NCA) solver for the impurity problem, which seems to partially conserve the energy \(E_{Mig}\), leading to a different value compared to \(E_{\chi}\).
Empirically, we found that by subtracting half of the kinetic energy (\(K/2\)) from \(E_{Mig}\), we achieve perfect agreement between \(E_{\chi}\) and \(E_{Mig}\). 
From this result one can conclude that the violation of energy conservation stems from the NCA approximation, which apparently
provides a more accurate result for the energy when obtained through the susceptibilities rather than from single-particle quantities.
Using higher-order impurity solvers to restore the missing piece of self-energy in \(E_{Mig}\) could cure the problem, but this approach increases the numerical expense and complicates efforts to reach long-time scales. A similar discrepancy between \(E_{\chi}\) and \(E_{Mig}\) has also been observed for Mott-insulators within DMFT.

\begin{figure}[t!]
\includegraphics[width=0.88\linewidth, trim={10pt 10pt 10pt 10pt}, clip]{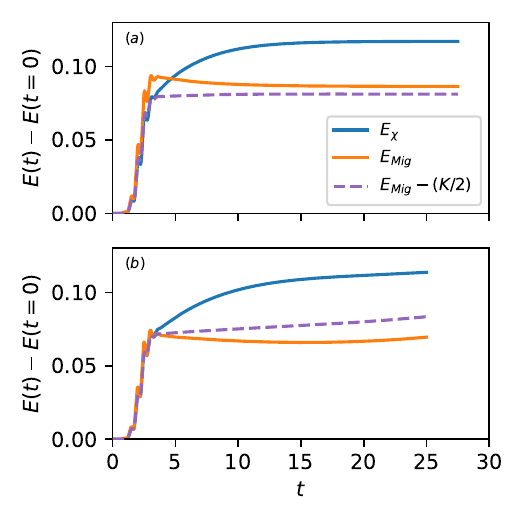}
\caption{The total energy change from the initial equilibrium state is plotted within D-$GW$ for metals and Mott-insulators in the upper and lower panels, respectively. 
\label{fig:total_energies_DGW}}
\end{figure}

A conserving approximation for the \mbox{D-$GW$} theory is possible, because in this approach the single (self-energy) and two-particle (polarization operator) quantities are also obtained from the functional ${\Phi[\tilde{G},\tilde{W}]}$ that corresponds to the partially bosonized dual action~\eqref{S_DB_fin}.
Since the transformation from the dual to the original space is exact, this should also provide a conserving description for the initial system~\cite{PhysRevB.79.045133}. 
However, there is an open question, which approach for calculating the energy should be used in this case.
Above, we have found that within the NCA framework the \(E_{\chi}\) approach to calculating the energy is more accurate.
However, on the contrary to DMFT, the self-energy in \mbox{D-$GW$} has a non-local contribution.
This means that the consistent calculation of the two-particle susceptibilities should account for the vertex corrections corresponding to the collective fluctuations accounted for in the self-energy.
This procedure is prohibitively expensive numerically, and in the \mbox{D-$GW$} approach, as well as in the other time-dependent diagrammatic methods, these corrections are neglected, and the polarization operator has a simple ``bubble'' form~\eqref{eq:Pi_dual_para}.
Therefore, the Galitskii-Migdal ($E_{Mig}$) way of calculating the energy through the self-energy is expected to be more accurate, but in this case we again encounter the issue that the NCA approximation misses some crucial contribution to the self-energy.
Within DMFT, this contribution is likely to be the half of the kinetic energy, as has been found above.
However, there is no grantee that the same contribution is missing in the \mbox{D-$GW$} approach due to a non-trivial relation between the lattice and dual quantities in the theory.

To investigate this, we examine total energy conservation, as illustrated in Fig.~\ref{fig:total_energies_DGW} for (a) metals and (b) Mott insulators. In contrast to DMFT, neither \(E_{\chi}\) nor \(E_{\text{Mig}}\) conserves energy immediately after the pulse. However, both approaches do conserve energy over longer time scales. 
By subtracting half of the kinetic energy from \(E_{\text{Mig}}\), we find that total energy conservation is maintained over an extended period in metals. 
In contrast, Mott insulators show a slight drift in long-term energy conservation (for ${t \geq 20}$) due to a drift in electron density, which can not be controlled with Matsubara and real-time grid sizes. Therefore, we cannot definitively determine whether the violation of energy conservation in \mbox{D-$GW$} is inherited from the NCA approximation or arises from the diagrammatic approximation, including the instantaneous approximation for the three-point vertex $\Lambda$.

\subsection{Relaxation dynamics of total energy components}

Energy relaxation in photo-excited metals and narrow-gap Mott insulators typically occurs through intraband (within the Hubbard bands) and interband (between lower and upper Hubbard bands) relaxation processes involving charge carriers. A common example of these processes includes electron-electron scattering, scattering of electrons with charge and spin degrees of freedom, and impact ionization phenomena \cite{Yuta2023}. The latter refers to an interband relaxation process distinguished by increased doublon density, even after a pulse. A high-energy doublon can create additional low-energy doublon-holon pairs through particle-particle scattering. In contrast, intraband electron-electron scattering and scattering involving low-energy degrees of freedom do not change the doublon density and primarily lead to thermalization.

In DMFT, the potential energy is directly proportional to the number of doublons, which relax according to various processes based on the initial photo-excited state. Since DMFT does not consider non-local fluctuations, the potential energy gain in the photo-excited state is compensated by the loss of electronic kinetic energy. Hence, in the pure electronic models, they both relax simultaneously. However, this may not be the case in \mbox{D-$GW$} theory, where non-local charge and spin degrees of freedom can affect their dynamics independently. 

\begin{figure*}[t!]
\includegraphics[width=1\linewidth]{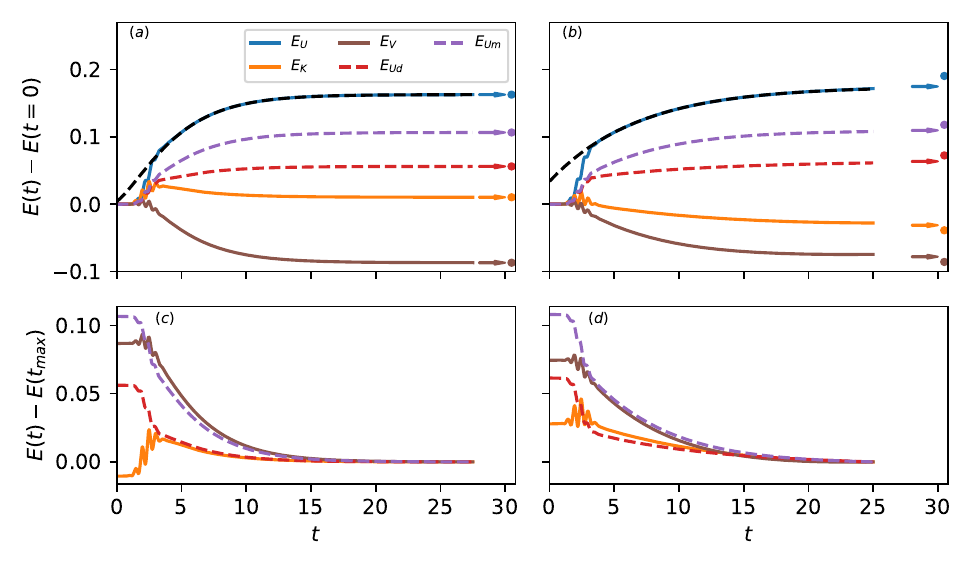}
\caption{The relaxation of total-energy components of the photo-doped metals(a) and Mott-insulators (b) is compared in \mbox{D-$GW$} for a given excitation protocol. The arrows are the residual value of exponential fit at  $t=\infty$ for a given energy component. The circles are the unique thermal equilibrium state values obtained from the total energy check. The black dashed line is the double (single) exponential fit of $E_{U}$ in metals (Mott-insulator). The change in total-energy components from their maximum simulated time value is plotted in panels (c) and (d) for metals and Mott-insulators, respectively. The red (violet) dashed line in the lower panel represents -$E_{Ud}$ (-$E_{Um}$).
\label{fig:energy_component_DGW}}
\end{figure*}

To address this question, we plot the total energy components of metals and Mott insulators obtained from \mbox{D-$GW$} in Fig.~\ref{fig:energy_component_DGW}\,(a) and (b), respectively. This analysis also illustrates how the system, subjected to the same excitation protocol but starting from different initial states, relaxes after the pulse. The initial observation is that all components of the total energy exhibit oscillations with a frequency of $\omega_p$ and are damped by the end of the pulse (for $t \sim 4$). For metals and Mott insulators, the local potential energy $E_U$, proportional to the number of doublons, increases during the pulse. This confirms that an electric pulse injects photo-excited charge carriers, known as doublons, into the system. In metals, we observe an exponential increase in $E_U$ after the pulse, followed by a saturation phase over a more extended period. This behavior is a smooth crossover of doublon relaxation from inter-band to intra-band processes at $t > 15$. In contrast, we do not observe such a crossover in the Mott insulators, which likely happens at much longer timescales ($t > 25$). 

On the other hand, the electronic kinetic energy $E_K$ in the metals also increases during the pulse, but the increase is less significant compared to $E_U$. This delocalization behavior can be attributed to the disappearance of quasi-particles due to the rise in the effective temperature of the system, which can be identified in the single-particle spectral functions. After the pulse, the kinetic energy slightly decays and eventually reaches a constant value over a longer period. However, in Mott insulators, the kinetic energy during the pulse oscillates around its initial value before starting to decay. In the presence of the Mott gap in the initial state, the electrons in the Brillouin zone experience an oscillating electric field. As a result, the kinetic energy of Mott insulators neither localizes nor delocalizes during the pulse. 

Creating a single photo-excited doublon state increases the local potential energy by Hubbard interaction \(U\). On the other hand, the non-local interaction $V$ energetically favors the creation of doublon-holon pairs at neighboring lattice sites rather than random ones. Such an excitation disrupts the non-local density-density interaction among neighboring lattice sites (utmost four sites), which usually compensates for the increase in local potential energy.
As expected, the non-local potential energy \(E_V\), as shown in Fig.~\ref{fig:energy_component_DGW}, decreases during the pulse for both metals and Mott insulators to offset the local potential energy increase. 
Additionally, the close-packed doublon-holon pattern favored by $V$ strongly suppresses magnetic excitations in this region.
Consequently, we find that \(E_V\) mirrors the dynamics of local potential energy related to spin excitations \(E_{Um}\) as illustrated in Fig.~\ref{fig:energy_component_DGW}\,(c) and (d).
On the other hand, the creation of local charge excitations blocks the electron momentum, so \(E_{Ud}\) tracks the dynamics of electronic kinetic energy \(E_K\) for both metals and Mott insulators [see Fig.~\ref{fig:energy_component_DGW}(c) and (d)]. 
Since the local charge and spin potential energy are proportional to the Hubbard interaction \(U\), the number of local spin excitations created in the transient state is nearly double that of the local charge excitations. 

\begin{figure*}[t!]
\includegraphics[width=\linewidth]{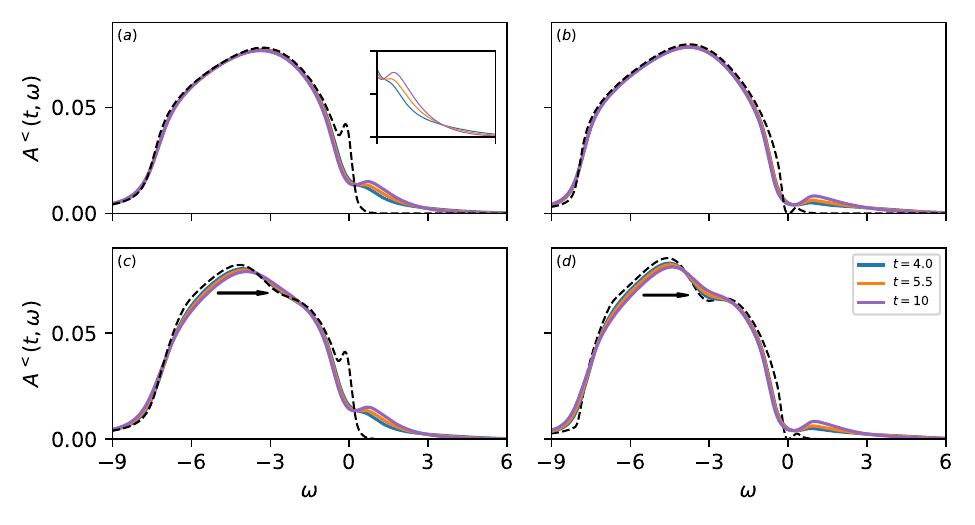}
\caption{The early (immediately after the pulse) time evolution of the occupied density of states for metals (left panels) and Mott-insulators (right panels) is compared between d-only (a,b) and \mbox{D-$GW$} (c,d). The inset zooms in on the evolution of the occupied spectrum above the 
Fermi level. The dashed black lines indicate the occupied density of states of the initial equilibrium state. The horizontal arrows indicate the direction of transfer of spectral weight at the lower Hubbard band in the \mbox{D-$GW$} density of states. 
\label{fig:spectrum_early}}
\end{figure*}

We extract the relaxation time scales of the total energy components using a function that characterizes relaxation dynamics over an extended time scale. In metals, we find that the relaxation dynamics can be well described by a double exponential function: $f(t) = a(t=\infty) + b \exp(-t/\tau_{h}) + c \exp(-t/\tau_{l})$. Fig.~\ref{fig:energy_component_DGW}\,(a) illustrates it for local potential energy. The relaxation dynamics involve two-time scales, denoted as \(\tau_{h}\) and \(\tau_{l}\). The microscopic origin of this two-time scale behavior stems from the interaction of a photo-excited high-energy doublon during the impact ionization process. As the high-energy doublon loses some of its kinetic energy, it creates an additional low-energy doublon-hole pair through particle-particle scattering, represented by the reaction: $\text{doublon}_{\text{high}} \rightarrow \text{doublon}_{\text{low}} + \text{doublon}_{\text{low}} + \text{holon}_{\text{low}}$. This particle-particle scattering mechanism accounts for the distinct relaxation time scales observed in the system. 

A similar pair-creation process can occur for holons in the photo-excited state. Due to particle-hole symmetry, the net effect of the impact ionization process is that each high-energy double-hole pair produces three low-energy double-hole pairs. Since the impact ionization process involves high-energy and low-energy doublons, one can expect a two-time scale behavior in the relaxation dynamics. These time scales are denoted by \(\tau_{h}\) and \(\tau_{l}\), corresponding to high-energy and low-energy doublons (holons), respectively. 

\begin{table}[t!]
\caption{Relaxation time scales of \mbox{D-$GW$} energy components}
\label{tab:DGW_taus}
\begin{ruledtabular}
\begin{tabular}{l  c  c  c  c  }
& \multicolumn{2}{c} {Metal} & 
\multicolumn{2}{c} {Mott-insulator} \\
\hline
   &  $\tau_{h}$ & $\tau_{l}$ & $\tau_{h}$  & $\tau_{l}$ \\
\hline \\
$E_K$ & 1.20 & 3.58 & 9.40 & - \\
$E_{Ud}$ & 1.23 & 3.33 & 9.10 & - \\
$E_V$ & 2.00 & 3.14 & 6.00 & - \\
$E_{Um}$ & 2.04 &  3.13 & 6.00 & - \\
$E_{U}$ & 2.02 & 3.14 & 6.80 & - \\
\end{tabular}
\end{ruledtabular}
\end{table}

In Mott insulators, a single exponential function ${f(t) = a(t=\infty) + b \exp(-t/\tau_{h})}$ is sufficient to explain the relaxation dynamics, as shown in Fig.~\ref{fig:energy_component_DGW}(b). We believe that the lack of long-term data prevents us from determining the time scale \(\tau_{l}\) associated with low-energy doublons (holons). We summarize these time scales for metals and Mott insulators in Table~\ref{tab:DGW_taus}. The relaxation times of all energy components in metals and Mott insulators are divided into two groups: one comprising \(E_K\) and \(E_{Ud}\), while the other includes the remaining three components. It is worth noting that the relaxation time scales \(\tau_{h}\) for Mott insulators are an order of magnitude longer than those for metals.  

\subsection{Time-dependent spectral functions}

\begin{figure*}[t!]
\includegraphics[width=\linewidth]{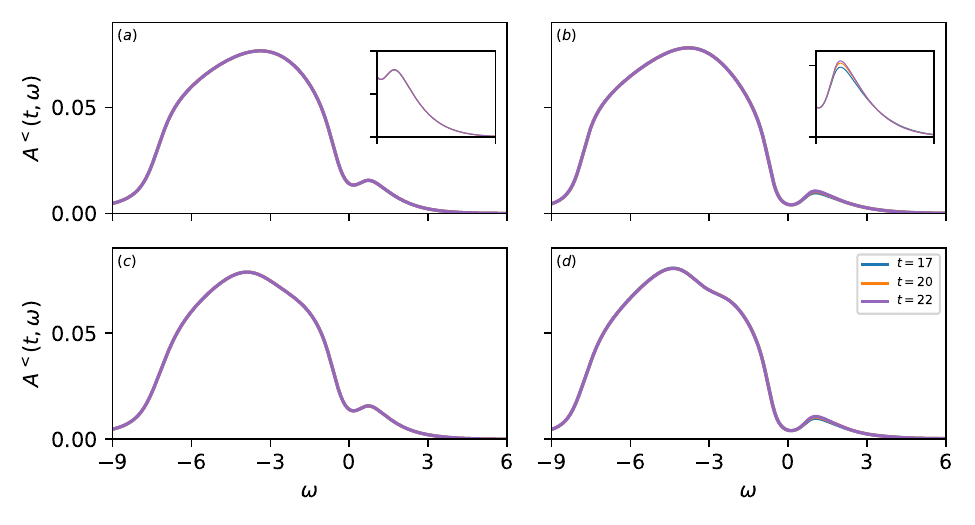}
\caption{The time evolution of the occupied density of states
at longer times ($t>15$) is compared between \mbox{D-$GW^{\rm d}$} (a,b) and \mbox{D-$GW$} (c,d) for metals (left panels) and Mott insulators (right panels). The inset shows the evolution of the occupied spectrum above the Fermi level. 
\label{fig:spectrum_lat}}
\end{figure*}

The integrated occupied density of states ${A^{<}(t,\omega)}$ above the Fermi level is closely related to the doublon number. Therefore, understanding their time evolution in the photo-excited state helps identify the intra- and interband relaxation regimes and their associated spectral features. In Fig.~\ref{fig:spectrum_early}, the time-resolved ${A^{<}(t,\omega)}$ for metals and Mott insulators is plotted immediately after the electric pulse (${t>4}$) for both \mbox{D-$GW$} and a diagrammatic method known as \mbox{D-$GW^{\rm d}$}. The latter includes only non-local charge fluctuations in the \mbox{D-$GW$} electronic self-energy while disregarding spin fluctuations. This is achieved by independently setting the renormalized interactions \(W^{d/m}\) in the charge (d) and spin (m) channels to zero in Eq.~\ref{eq:Sigma_dual_para}. This approach allows us to study the effects of spin fluctuations \mbox{D-$GW^{\rm m}$} or charge fluctuations \mbox{D-$GW^{\rm d}$} on spectral functions separately. 

Immediately after the pulse, the photo-excited charge carriers in metals and Mott insulators relax primarily through an impact ionization phenomenon (see Fig.~\ref{fig:energy_component_DGW}). In the initial equilibrium state, the occupied density of states below the Fermi level—the lower Hubbard band—is filled. In contrast, the upper Hubbard band above the Fermi level is nearly unoccupied. A prominent quasi-particle peak at the Fermi level distinguishes a metallic state [Fig.~\ref{fig:spectrum_early}(a) and (c)] from a Mott insulator [Fig.~\ref{fig:spectrum_early}(b) and (d)]. The applied electric pulse causes part of the spectrum to move from the lower to the upper Hubbard band. The low-energy quasi-particle peak completely melts in metals, whereas the gap fills up in Mott insulators. The time evolution of the photo-excited state after the pulse is similar for both metals and Mott insulators. 

A key characteristic of the impact ionization process in the occupied density of states is the transfer of spectral weight. This transfer occurs from the upper edge of the upper Hubbard band (approximately at ${\omega \sim 4}$) to the lower edge of the upper Hubbard band (around ${\omega \sim 1}$). This behavior is illustrated in the inset of Fig.~\ref{fig:spectrum_early}. Additionally, the ratio of spectral gain (integrated spectrum) at ${\omega_{gain} \sim 1}$ is nearly three times the spectral loss at ${\omega_{loss} \sim 4}$. This observation confirms that nearly three low-energy doublon-holon pairs are produced from a single high-energy doublon-holon pair \cite{werner2014}. Notably, in the \mbox{D-$GW$} calculations, aside from the effect of impact ionization, spectral weight is transferred from the upper edge of the lower Hubbard band to its lower edge. The feature indicated by the arrow in the lower Hubbard band, as shown in Fig.~\ref{fig:spectrum_early}(c) and (d), which is absent in \mbox{D-$GW^{\rm d}$} states, can be attributed to a relaxation process involving magnetic fluctuations.

We further examine the time evolution of ${A^{<}(t,\omega)}$ over a longer duration, where the intraband relaxation of charge carriers dominates. In the case of metals, the ${A^{<}(t,\omega)}$ obtained from \mbox{D-$GW$} and \mbox{D-$GW^{\rm d}$} approaches, as shown in Fig.~\ref{fig:spectrum_lat} (a) and (c), does not change further over time. The constant distribution of spectral weight at the upper Hubbard band (see the inset of Fig.~\ref{fig:spectrum_lat} (a)) indicates that the intraband relaxation does not alter the number of doublons. In Mott insulators, since the time scales for intraband relaxation have not yet been reached, a small fraction of spectral weight continues to transfer from high-energy to low-energy states above the Fermi level, as illustrated in the inset of Fig.~\ref{fig:spectrum_lat} (b). Interestingly, a relaxation process involving magnetic fluctuations at the lower Hubbard band in D-GW is absent over longer periods. They are only active immediately after the pulse (interband relaxation regime) [see Fig.~\ref{fig:spectrum_early}(c) and (d)]. 

In small cluster calculations, it has been observed that antiferromagnetic fluctuations do not favor impact ionization \cite{kauch2020}. This is because a high-energy doublon transfers its excess kinetic energy to the underlying spin background instead of generating an additional doublon-holon pair. However, the situation is less clear for extended systems. The time evolution of the occupied density of states in \mbox{D-$GW$} indicates that impact ionization remains favorable even in magnetic fluctuations. We identified signs of energy transfer to the underlying antiferromagnetic spin fluctuations during impact ionization. This is evidenced by spectral weight transfer at the edges of the lower Hubbard band immediately after the pulse. This phenomenon can be measured using time-resolved photoemission spectroscopy. It has been demonstrated through equilibrium studies that the kinks at the lower edges of the Hubbard band are linked to electromagnetic coupling, as evidenced by various theoretical approaches \cite{PhysRevX.10.041023, PhysRevX.10.041023,wang2020emergence,PhysRevLett.108.076401,PhysRevB.92.075119, Silke2025}. We have observed this feature in the equilibrium spectra presented in Fig.~\ref{fig:optics}(b).

\subsection{Thermalization of photo-excited states}

According to the eigenstate thermalization hypothesis, a closed quantum many-body system driven out of equilibrium will eventually reach thermalization over a longer period~\cite{Mark1994, Deutsch_2018}. However, the evolution of a photo-excited state with non-local collective electronic fluctuations toward thermalization is not yet fully understood. We investigate thermalization by comparing the properties of a time-dependent state to a uniquely defined thermal equilibrium state. This thermal equilibrium state is determined by the condition that its energy matches the energy of a photo-excited state, given by the equation $E(t>4) = \frac{\text{Tr}[e^{-\mathcal{H}/T_{\text{eff}}} \mathcal{H}]}{\text{Tr}[e^{-\mathcal{H}/T_{eff}}]}$. This procedure results in a thermal equilibrium state with ${T_{\text{eff}} = 0.85}$ for metals and 0.95 for Mott insulators. The components of total energy obtained from these thermal equilibrium states are plotted in Fig.~\ref{fig:energy_component_DGW} for both metals and Mott insulators. The perfect agreement of the residual energy components $a(t=\infty)$ with the thermal equilibrium state confirms the thermalization of the photo-excited metals, as shown in Fig.~\ref{fig:energy_component_DGW}(a). In contrast, the discrepancy between the thermal values and the photo-excited state highlights the pre-thermal nature of the transient state for Mott insulators, as observed in DMFT\cite{werner2014}.

\begin{figure}[t!]
\includegraphics[width=0.9\linewidth, trim={10pt 10pt 10pt 10pt}, clip]{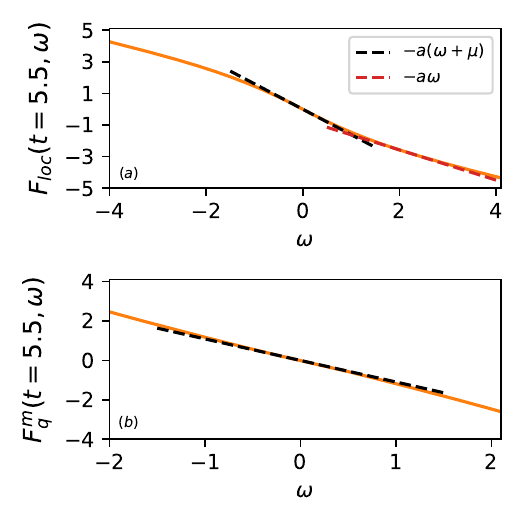}
\caption{The function related to the fluctuation-dissipation theorem is plotted for local electronic spectrum  (a) and ferromagnetic fluctuations at $\Gamma$ point in the 2d square Brillouin zone. The dashed lines are the linear fits of $F_{loc}(t,\omega)$ and $F^{m}_{q=\Gamma}(t,\omega)$.
\label{fig:F_D}}
\end{figure}

\begin{figure}[t!]
\includegraphics[width=0.95\linewidth, trim={10pt 10pt 10pt 10pt}, clip]{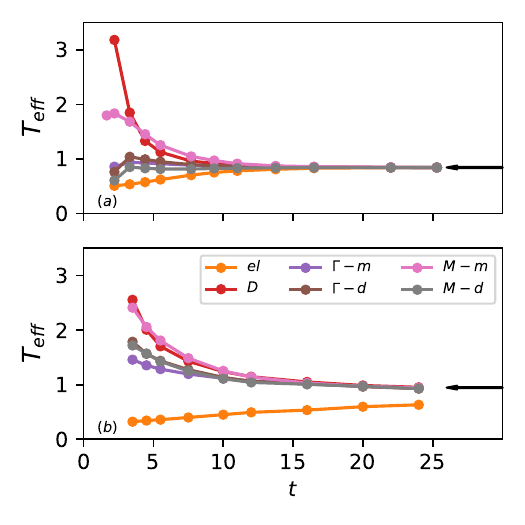}
\caption{The time-dependent effective temperature of electrons ($el$), doublons ($D$), charge ($d$) and spin fluctuations ($m$) are obtained from the slope of a linear fit to $F_{loc}(t,\omega)$, $F^{d/m}_{q}(t,\omega)$ for metals (a) and Mott-insulators (b). The charge and spin fluctuations extracted at two different $q=(\Gamma, M)$ points in the 2d square Brillouin zone. The black arrow is the effective temperature of the unique thermal state obtained from the total energy check. 
\label{fig:T_eff}}
\end{figure}

The total energy procedure determines whether the photo-excited electronic state reaches thermal equilibrium over time. To understand how and when a system with non-local fluctuations attains a thermal state, we analyzed the dynamics of the effective temperature for electrons and charge and spin fluctuations near the Fermi level (with $\omega \sim 0$). In Fig.~\ref{fig:F_D}, we present the distribution function for the local electronic spectral function $F_{loc}(t,\omega)$ and spin susceptibility at the ferromagnetic wave vector $F^{m}_{q=\Gamma}(t,\omega)$ defined in Eq.~\eqref{hkjwdlnsks}.
The electronic distribution function $F_{loc}(t,\omega)$ in Fig.~\ref{fig:F_D}(a) shows linear behavior around the Fermi level and transitions to a shifted linear behavior at high frequencies (between $\omega \sim 1.5$ and $4$). This latter behavior corresponds to the frequency range of spectral weight transfer in $A^{<}(t,\omega)$ at the upper Hubbard band, as shown in Fig.~\ref{fig:spectrum_early}. As mentioned earlier, the total spectral weight in this frequency range indicates the doublon number. Therefore, fitting a (shifted) linear function at the Fermi level allows us to determine the effective temperature of the (doublon) quasi-particles. Due to symmetric conditions, holons have the same temperature as doublons. This can be similarly derived from a linear fit of $F_{loc}(t,\omega)$ below the Fermi level. In Fig.~\ref{fig:F_D}(b), a linear fit of ${F^{m}_{q=\Gamma}(t,\omega)}$ 
near $\omega=0$ yields the effective temperature of ferromagnetic fluctuations at the specified time scale. The same procedure has also been applied to charge fluctuations at the \(\Gamma\) and $M$ points, where the latter represents the antiferro order in a given channel. The corresponding effective temperatures for metals and Mott insulators are shown in Fig.~\ref{fig:T_eff}(a) and (b), respectively.

On the otherhand, the electric pulse injects a finite amount of energy into the photo-excited system, distributed among quasi-particles, doublons, charge carriers, and spin fluctuations. This process raises their temperature from the initial equilibrium temperature of ${T = 1/\beta = 0.16}$. The increase in electronic kinetic energy within metals (see Fig.~\ref{fig:energy_component_DGW}(a)) during the electric pulse is attributed to a rise in the quasi-particle temperature, as illustrated in Fig.~\ref{fig:T_eff}(a). This increase in temperature, in turn, leads to a suppression of the quasi-particle lifetime. This effect is also reflected in the time evolution of the occupied density of states near the Fermi level (see Fig.~\ref{fig:spectrum_early}(a) and (c)), where the spectral weight is significantly diminished during the pulse.  

In addition to the dynamics of quasi-particles, an electric pulse triggers the creation of photo-excited doublons and anti-ferromagnetic (AFM) fluctuations, which rapidly increases their temperature. In metals (see Fig.~\ref{fig:T_eff}(a)), other collective fluctuations, such as charge and ferromagnetic (FM), do not respond similarly. This discrepancy is primarily due to itinerant electrons, which significantly contribute to AFM fluctuations in comparison to FM fluctuations. In Mott insulators, electronic correlations are primarily local, independent of momentum. As a result, an electric pulse activates all collective dynamics, with doublons and AFM fluctuations remaining the most prominent, as illustrated in Fig.~\ref{fig:T_eff}(b). 

Over time, the doublons and AFM fluctuations in metals decrease temperature by effectively transferring their energy to the quasi-particle excitations. This process further raises the temperature of the damped quasi-particles. This energy transfer is nearly complete by the time scale of approximately ${t \sim 17}$, and all fluctuations—including charge and ferromagnetic (FM) —equilibrate at the same temperature. It has been illustrated in Fig.~\ref{fig:T_eff}(a). The perfect alignment of this temperature with the unique thermal state (indicated by the black arrow in Fig.~\ref{fig:T_eff}(a)) around ${t \sim 17}$ confirms that this is the typical time scale for thermalization in metals.

In Mott insulators (see Fig. \ref{fig:T_eff}(b)), all collective fluctuations transfer their energy to quasi-particle excitations as the temperature decreases, eventually reaching a specific temperature around ${t \sim 20}$. This temperature corresponds to a distinct thermal state (indicated by the black arrow), but it does not align with the temperature of the quasi-particles. In the Mott state, there are initially no quasi-particles present. However, when an electric pulse is applied, it shifts the spectral weight from the lower to the upper Hubbard band, effectively filling the Mott gap (see Fig. \ref{fig:spectrum_early} and \ref{fig:spectrum_lat}).

The finite spectral weight at the Fermi level in the transient state indicates a behavior similar to that of a bad-metal, characterized by highly damped quasi-particles. An ineffective energy transfer between collective fluctuations and quasi-particle excitations results in a temperature mismatch, which slows down the thermalization process in Mott insulators. Thermalization occurs when the temperatures of the collective fluctuations and quasi-particles become equal. The inability to reach a single temperature state suggests the system does not entirely lose its initial correlations within the simulated time frame. As a result, much longer durations are needed to observe thermalization in Mott insulators.

\section{Conclusions\label{sec:conclusion}}

In this paper, we implemented a time-dependent \mbox{D-$GW$} approach on the L-shaped Keldysh contour. This method goes beyond one of the most advanced non-equilibrium approaches to date, namely the time-dependent $GW$+EDMFT method, by incorporating spatial magnetic fluctuations into the theory.
This self-consistent theoretical framework not only allows one to track the time evolution of the entire photo-excited system but also enables a microscopic analysis of the complex energy transfer among its subsystems, involving quasi-particles, doublons, charge carriers, and magnetic fluctuations after the applied pulse.
Additionally, we demonstrate that the \mbox{D-$GW$} approach conserves total energy in the strongly corrected metallic regime in the vicinity of the Mott transition. 
This is one of the key advantages of the developed method over other diagrammatic methods, such as time-dependent FLEX and TPSC+DMFT approaches, that cannot access this physical regime.

Using the \mbox{D-$GW$} approach, we study the dynamics of the half-filled single-band extended Hubbard model on a 2D square lattice, which displays strong collective charge and spin fluctuations. We excite this model with an ultrashort laser pulse near the metal-to-Mott-insulator transition. We aim to understand how and when the correlated metals and narrow-gap Mott insulators relax under the light pulse and achieve thermalization along with the non-local collective fluctuations. 

The energy relaxation dynamics of correlated metals are well described by a double exponential function, which originates from the impact ionization process - creating three low-energy doublon-holon pairs from a single high-energy doublon-holon pair through particle-particle scattering involving two-time scales. 
Such an impact ionization process in Mott insulators is evident in the dynamics of the local potential energy. We used a single exponential function to analyze the relaxation dynamics of high-energy doublons, which are significantly larger—by an order of magnitude—than those found in metals. Unfortunately, we could not determine the relaxation time scales of low-energy doublons due to the absence of longer time scales that were inaccessible in our simulation data.   The extended relaxation time scales of high-energy doublons highlight the slower thermalization processes in Mott insulators compared to metals, even when the same excitation protocol is applied.

The time evolution of the occupied density of states demonstrates a spectral weight transfer from the lower to the upper Hubbard band when an electric pulse is applied. This process also shows that the electric pulse disrupts the low-energy peak near the Fermi level, which is associated with quasi-particle excitations in metals, and fills the gap in Mott insulators. After the pulse concludes, the transient states in both scenarios resemble a bad-metal-like behavior and are indistinguishable. Since the integrated occupied density of states above the Fermi level relates closely to the number of doublons, we observe signatures of impact ionization in the dynamics of occupied density of states. For instance, there is a decrease in spectral weight, denoted as \(\omega_{loss}\), at the upper edge of the upper Hubbard band, and an increase in spectral weight, denoted as \(\omega_{gain}\), at the lower edge of the upper Hubbard band. Additionally, the ratio of spectral gain at \(\omega_{gain}\) is nearly three times greater than the spectral loss at \(\omega_{loss}\), indicating that approximately three low-energy doublon-hole pairs are created from a single high-energy doublon pair.

A noteworthy characteristic identified in the \mbox{D-$GW$} approach, which is absent in the \mbox{D-$GW^{d}$}, is the transfer of spectral weight from the lower edge of the lower Hubbard band to the upper edge of the lower Hubbard band. This spectral transfer occurs solely during the interband relaxation ($t \leq 10$) regime and is not observed in the intraband relaxation regime ($t>17$). Since the kinks at the edges of the Hubbard band in the spectral functions are linked to electron-magnetic coupling, this transfer at the edges of the lower Hubbard band facilitates the transfer of excess kinetic energy from high-energy doublons to the underlying antiferromagnetic (AF) spin fluctuations during the impact ionization process. The evolution of the occupied density of states derived from \mbox{D-$GW$} suggests a preference for impact ionization even in AF spin fluctuations in extended systems. These new features can be measured using time-resolved photoemission spectroscopy.     

We analyzed the evolution of the effective temperature of quasi-particles, doublons, and both charge and spin degrees of freedom. Our focus was on the transient state over an extended time and the timescale at which this transient state reaches thermalization. During the pulse, the itinerant nature of magnetic moments in metals causes the temperature of photo-excited doublons to be higher than that of AFM fluctuations, while other collective fluctuations have a minimal impact. Over time, doublons and AFM fluctuations in metals lower their temperature by efficiently transferring energy to quasiparticle excitations, increasing the quasiparticles' temperature further. This energy transfer process concludes around the time scale ${t \sim 17}$, at which point all degrees of freedom converge to a single thermal state temperature. This marks the thermalization of correlated metals.

Narrow-gap Mott insulators are found to behave different from a metal. Due to the localized nature of magnetic moments, an electric pulse transfers excitation energy to all collective fluctuations, with doublons and AFM being the most prominent. Over time, the temperature of these collective fluctuations decreases as energy is transferred to quasi-particle excitations. They eventually stabilize at a constant temperature around ${t \sim 20}$, which corresponds to a unique thermal state temperature. This energy transfer raises the temperature of the quasi-particles but does not lead to alignment with the collective fluctuations. Quasiparticle excitations are not present in the initial equilibrium state. However, when an electric pulse is applied, it fills the Mott gap and introduces finite spectral weight near the Fermi level. This transient state resembles a ``bad metal," characterized by damped quasiparticles that gradually acquire an effective temperature over time. 

Inefficient energy transfer between collective fluctuations and quasiparticle excitations results in a temperature mismatch, leading to a slowdown of thermalization in Mott insulators. This behavior suggests that the transient state is pre-thermal and ultimately progresses to thermalization, where all degrees of freedom achieve an equal temperature. The delay in thermalization within Mott insulators indicates that the system retains its initial state correlations during the simulated time, even though the transient state displays characteristics similar to those of a bad-metal.    
 
\mbox{D-$GW$} has several potential future applications. First, we aim to extend it to multi-orbital lattice models and investigate how non-local collective fluctuations influence Hund's and crystal-field-driven electron dynamics. The truncation of memory kernels in Dynamical Mean-Field Theory (DMFT) enables us to reach thermalization time scales in Mott insulators \cite{Stahl2022}. Applying this method within a \mbox{D-$GW$} framework could uncover non-thermal states resulting from competing collective electronic fluctuations. Currently, in the \mbox{D-$GW$} setup, we determine the impurity and dual Green's functions in a self-consistent manner but have not accounted for the impact of non-local fluctuations on the impurity Green's functions. Establishing a complete self-consistency procedure is another interesting direction for future research. Another approach could involve calculating the local three-particle vertex using real-time exact diagonalization techniques on small clusters, then incorporating these results into the non-local self-energy while solving the reference problem using lowest-order strong coupling impurity solvers. Additionally, a direct application of \mbox{D-TRILEX} to non-equilibrium steady-state problems is immediately viable, thanks to the recent tensor train implementation of high-order strong coupling impurity models 
\cite{eckstein2024,Kim2025TCI}.

\begin{acknowledgments}
N.D., H.U.R.S. and A.I.L. acknowledge support from the European Research Council via Synergy Grant No. 854843 (the FASTCORR project). H.U.R.S. acknowledges financial support from the Swedish Research Council (Vetenskapsrådet, VR), grant number 2024-04652. 
M.E. and  A.I.L. acknowledge support from the Deutsche Forschungs-gemeinschaft 
through the research unit QUAST, FOR 5249, Project ID No. 449872909.
M. E. is supported by the Cluster of Excellence ”CUI: Advanced Imaging of Matter“ of the Deutsche Forschungsgemeinschaft (DFG) – EXC 2056 – project ID 390715994.
E.A.S. acknowledges support from LabEx PALM Paris-Saclay through the CEBULI project and also from CNRS through the Physique Tremplin project UFEX. This research was supported in part through the Maxwell computational resources operated at Deutsches Elektronen-Synchrotron DESY, Hamburg, Germany.
\end{acknowledgments}

\bibliography{apssamp}

\onecolumngrid
\appendix
\section{}
\subsection{DMFT on Keldysh contour:}
\label{app:DMFT}
In the DMFT approximation, the lattice Green's function in the paramagnetic case 
\begin{align}
        {G}^{-1,DMFT}_{k\sigma}(z,z') = [i\partial_z + \mu - \epsilon_k(z]\delta_{c}(z,z') - \Sigma_{\rm imp}^{\sigma}(z,z')
        \label{eq:DMFT_G}
\end{align}
is related to the impurity Green's function $g_{\sigma}(z,z')$ through $\sum_k G_{k\sigma}^{DMFT}(z,z') = g_{\sigma}(z,z')$. The impurity propagator
is obtained from the solution of an effective quantum impurity model with local action
\begin{align}
        {\cal S}_{\rm imp} = \int_{\cal C} dz \, dz' \Bigg\{\sum_{\sigma} c^{*}_{\sigma}(z) \left[ (i \partial_z + \mu - \epsilon_{loc}(z))\delta_{\cal C}(z,z') - \Delta_{\sigma}(z,z')\right] c_{\sigma} (z') - U \delta_{\cal C}(z,z') n_{\uparrow}(z) n_{\downarrow}(z') \Bigg\}
\end{align}
Here we have separated the singular part of the hybridization function $\epsilon_{loc}(z)= \sum_{k}\epsilon_k (z)$ from the regular contribution of $\Delta_{\sigma}(z,z')$.
The auxiliary impurity problem fulfils the following Dyson equation
\begin{equation}
        g^{-1}_{\sigma}(z,z') = (i\partial_z + \mu -\epsilon_{loc}(z))\delta_{\cal C}(z,z') - \Delta_{\sigma}(z,z') - \Sigma_{\rm imp}^{\sigma}(z,z')
        \label{eq:NCA_G}
\end{equation}
To solve DMFT equations on a square lattice, we introduce the isolated impurity Green’s function $g_0$, which is defined via the impurity Dyson
equation for $\Delta_{\sigma} = 0$,
\begin{align}
g^{-1}_{0,\sigma}(z,z') = (i\partial_z + \mu -\epsilon_{loc}(z))\delta_{\cal C}(z,z') - \Sigma_{\rm imp}^{\sigma}(z,z')
\end{align}
In order to compute $g$ one can reformulate equation~\ref{eq:NCA_G} in an integral form,
\begin{align}
        g_{\sigma}(z,z') = g_{0,\sigma}(z,z') + \int_{\cal C} dz''\,dz'''\, g_{0,\sigma}(z,z'') \, \Delta_{\sigma}(z'',z''') \, g_{\sigma}(z''',z')
\label{eq:g0}
\end{align}
Next one computes momentum-resolved DMFT Green’s function ${G}_{k\sigma}^{DMFT}$ from the Dyson equation~\ref{eq:DMFT_G} in the momentum representation,
\begin{align}
        {G}^{-1,DMFT}_{k\sigma}(z,z') = g^{-1}_{0,\sigma}(z,z') - \left[\epsilon_{k}(z)-\epsilon_{loc}(z\right]\delta_{\cal C}(z,z')
\end{align}
It reads in the integral form
\begin{align}
        {G}_{k\sigma}^{DMFT}(z,z') = g_{0,\sigma}(z,z') + \int_{\cal C} dz''\, g_{0,\sigma}(z,z'') \left[\epsilon_{k}(z'')-\epsilon_{loc}(z'')\right] \check{G}_{k\sigma}(z'',z')
\label{eq:GD}
\end{align}
To compute the updated $\Delta_{\sigma}$ we start by summing above equation over momentum $\bf{k}$
\begin{align}
        \sum_k {G}_{k\sigma}^{DMFT}(z,z') &= g_{0,\sigma}(z,z') + \int_{\cal C} dz''\, g_{0,\sigma}(z,z'') \left\{\sum_{k} \left[\epsilon_{k}(z'')-\epsilon_{loc}(z'')\right] {G}_{k\sigma}^{DMFT}(z'',z') \right\}\\
        g_{\sigma}(z,z') &= g_{0,\sigma}(z,z') + \int_{\cal C} dz''\,dz'''\, g_{0,\sigma}(z,z'') \, \Delta_{\sigma}(z'',z''') \, g_{\sigma}(z''',z')
\end{align}
Using the DMFT self-consistency condition, we get
\begin{align}
        g_{1,\sigma}(z,z') = \sum_{k} \left[\epsilon_{k}(z)-\epsilon_{loc}(z\right] {G}_{k\sigma}^{DMFT}(z,z') = \int_{\cal C} dz''\, \Delta_{\sigma}(z,z'') \, g_{\sigma}(z'',z')
\end{align}
After inserting the conjugate of the equation~\ref{eq:g0} and \ref{eq:GD} at l.h.s and r.h.s in the above equation, one finds

\begin{align}
        &\sum_k \left[\epsilon_{k}(z)-\epsilon_{loc}(z)\right] \left[g_{0,\sigma}(z,z') +  \int_{\cal C} dz''\, {G}_{k\sigma}^{DMFT}(z,z'') \left[\epsilon_{k}(z'')-\epsilon_{loc}(z'')\right] g_{0,\sigma}(z'',z') \right] = \notag\\
        &\int_{\cal C} dz''\, \Delta_{\sigma}(z,z'') \left[ g_{0,\sigma}(z'',z') + \int_{\cal C} dz'''\,dz''''\, g_{\sigma}(z'',z''') \, \Delta_{\sigma}(z''',z'''') \, g_{0,\sigma}(z'''',z') \right]
\end{align}
which results in
\begin{align}
g_{2,\sigma}(z,z')
&= \sum_{k} \left[\epsilon_{k}(z)-\epsilon_{loc}(z)\right]\delta_{\cal C}(z,z') + \left[\epsilon_{k}(z)-\epsilon_{loc}(z)\right] {G}_{k\sigma}^{DMFT}(z,z') \left[\epsilon_{k}(z'')-\epsilon_{loc}(z'')\right] \notag\\
        &= \sum_{k} \left[\epsilon_{k}(z)-\epsilon_{loc}(z)\right] {G}_{k\sigma}^{DMFT}(z,z') \left[\epsilon_{k}(z'')-\epsilon_{loc}(z'')\right] \notag\\
        &= \Delta_{\sigma}(z,z') + \int_{\cal C} dz''\,dz'''\, \Delta_{\sigma}(z,z'') \, g_{\sigma}(z'',z''') \, \Delta_{\sigma}(z''',z')
\end{align}
or
\begin{align}
g_{2,\sigma}(z,z') =
        \int_{\cal C} dz'' \left[\delta_{\cal C}(z,z'') + g_{1,\sigma}(z,z'') \right] \Delta_{\sigma}(z'',z')
\label{eq:Delta_upd}
\end{align}
The solution of impurity polarization $\Pi^{\varsigma}_{\rm imp}$ in equation~\ref{eq:imp_w} can be obtained from the impurity susceptibility as
\begin{align}
\int_{\cal C} dz'' \, \left[\delta_{\cal C}(z,z'') + \chi^{\varsigma}(z,z'') U^{\varsigma}\right]\Pi^{\varsigma}_{\rm imp}(z'',z') = \chi^{\varsigma}(z,z')
\label{eq:Pi_imp}
\end{align}

\end{document}